\documentclass[review,3p,twocolumn]{elsarticle}

\usepackage[hidelinks]{hyperref}

\usepackage{booktabs}       
\usepackage{pgfplots}       
\usepackage{pgfplotstable}  
\usepackage{tikz}           
\usetikzlibrary{patterns}
\usepackage{subcaption}

\makeatletter
\def\ps@pprintTitle{%
   \let\@oddhead\@empty
   \let\@evenhead\@empty
   \let\@oddfoot\@empty
   \let\@evenfoot\@oddfoot
}
\makeatother

\usepackage{listings}
\definecolor{strings_color}{rgb}{0.49,0.49,0.49}
\definecolor{comments_color}{rgb}{0.49,0.49,0.49}
\definecolor{border_color}{rgb}{0.81,0.81,0.81}
\usepackage{caption}
\usepackage{chngcntr} 

\journal{Parallel Computing Journal}

\begin{document}

\begin{frontmatter}

\title{MPI Windows on Storage for HPC Applications}





\author[kth]{Sergio~Rivas-Gomez}
\ead{sergiorg@kth.se}
\author[oak]{Roberto~Gioiosa}
\ead{gioiosar@ornl.gov}
\author[oak]{Ivy~Bo~Peng}
\ead{pengb@ornl.gov}
\author[oak]{Gokcen~Kestor}
\ead{kestorg@ornl.gov}
\author[sea]{Sai~Narasimhamurthy}
\ead{sai.narasimhamurthy@seagate.com}
\author[kth]{Erwin~Laure}
\ead{erwinl@kth.se}
\author[kth]{Stefano~Markidis}
\ead{markidis@kth.se}
\address[kth]{KTH Royal Institute of Technology, Stockholm 10044, Sweden}
\address[oak]{Oak Ridge National Laboratory, Oak Ridge, TN 37830, USA}
\address[sea]{Seagate Systems UK, Havant PO9 1SA, UK}

\begin{abstract}
Upcoming HPC clusters will feature hybrid memories and storage devices per compute node. In this work, we propose to use the MPI one-sided communication model and MPI windows as unique interface for programming memory and storage. We describe the design and implementation of MPI \emph{storage} windows, and present its benefits for out-of-core execution, parallel I/O and fault-tolerance. In addition, we explore the integration of heterogeneous window allocations, where memory and storage share a unified virtual address space. When performing large, irregular memory operations, we verify that MPI windows on local storage incurs a 55\% performance penalty on average. When using a Lustre parallel file system, ``asymmetric'' performance is observed with over 90\% degradation in writing operations. Nonetheless, experimental results of a Distributed Hash Table, the HACC I/O kernel mini-application, and a novel MapReduce implementation based on the use of MPI one-sided communication, indicate that the overall penalty of MPI windows on storage can be negligible in most cases in real-world applications.
\end{abstract}


\end{frontmatter}


\section{Introduction}
\label{1_Introduction}

Emerging storage technologies are evolving so rapidly that the existing gap between main memory and I/O subsystem performances is thinning \cite{nanavati2015non}. The new non-volatile solid-state technologies, such as flash, phase-change and spin-transfer torque memories \cite{caulfield2010understanding} provide bandwidth and latency close to those of DRAM memories. For this reason, memory and storage technologies are converging and storage will soon be seen as an extension of memory. Because of these new technological improvements, next-generation supercomputers will feature a variety of Non-Volatile RAM (NVRAM), with different performance characteristics and asymmetric read / write bandwidths, next to traditional hard disks and conventional DRAM~\cite{peng2016exploring, peng2017exploring}. In such systems, allocating and moving data often require the use of different programming interfaces to program separately memory and storage. For instance, MPI provides one sided-communication to access shared, intra-node memory, and distributed, inter-node memory. On the other hand, the MPI I/O interface separately provides support to read and write from files on storage. 

\begin{figure*}
  \begin{center}
    \includegraphics[width=1.81\columnwidth]{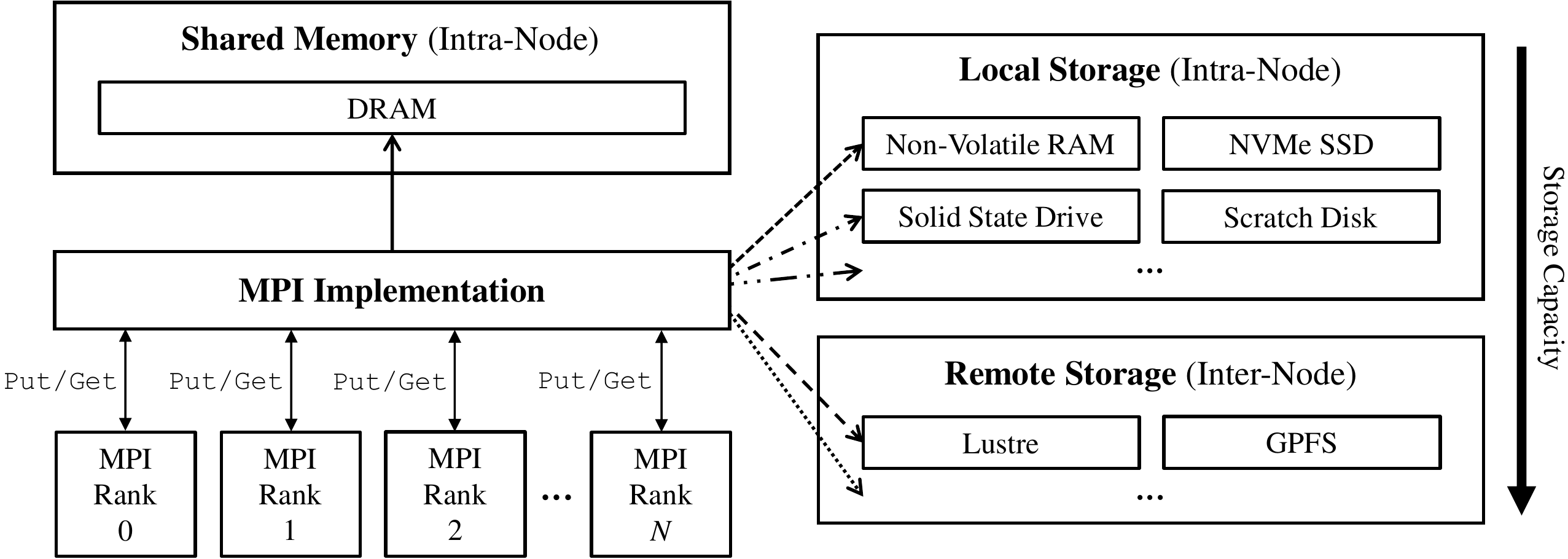}
    \vspace{0.21cm}
    \caption{With MPI storage windows, MPI implementors can enable HPC applications to seamlessly access different types of storage technologies, while maintaining a unified interface.}
    \label{fig:mpi_heterogeneous}
  \end{center}
\end{figure*}

In the same way that storage will seamlessly extend memory in the near future, programming interfaces for memory operations will also become interfaces for I/O operations. In this work, we aim at raising the level of programming abstraction by proposing the use of MPI one-sided communication and MPI windows as a unified interface to any of the available memory and storage technologies. MPI windows provide a familiar interface that can be used to program data movement among hybrid memory and storage subsystems (\autoref{fig:mpi_heterogeneous}). 
Simple \texttt{put} / \texttt{get} operations can be used for accessing local or remote windows. Support for shared memory programming using MPI windows is also possible, where these operations can be replaced by simple \texttt{store} / \texttt{load} memory operations \cite{gropp2014using, hoefler2013mpi}. In addition, we foresee the potential of \emph{heterogeneous} allocations that include memory and storage using a single virtual address space.

We design and implement MPI windows on storage, and evaluate its performance using two different testbeds. First, we use a single node with local hard disk and SSD, that serves us to mimic future computing nodes with memory and local storage. Additionally, we use a cluster that mounts a distributed file system, as in the majority of the current supercomputers. By evaluating our implementation with the Intel MPI RMA Benchmarks, we demonstrate that MPI windows on storage shows a negligible performance overhead for small data transfers compared to MPI windows in memory, when no storage synchronization is enforced. When performing large amounts of consecutive memory operations and enforcing data synchronization with storage, the penalty of MPI windows on local solid-state drives of a single computing node is approximately 55\% on average when compared to the performance of MPI windows in memory. When using a Lustre parallel file system, we observe an ``asymmetric'' performance of \texttt{put} and \texttt{get} operations, with over 90\% degradation in \texttt{put} operations.

Despite these limitations, MPI windows on storage provide benefits for HPC applications by enabling transparent out-of-core execution. The approach also defines a simple interface for performing I/O operations with MPI windows, as an alternative to POSIX I/O or MPI I/O. Furthermore, it enables the definition of novel use-cases, such as transparent fault-tolerance. Big Data applications and data analytics workloads are ideal candidates for using MPI windows on storage, as they process large amounts of data, stress the I/O subsystem with millions of \texttt{read} / \texttt{write} operations, and require support for checkpoint-restart mechanisms~\cite{reed2015exascale}. In this paper, we describe how MPI windows can be effectively used in this kind of applications and we show the performance of the approach using a Distributed Hash Table, the HACC I/O kernel, and a novel MapReduce implementation based on the use of MPI one-sided communication. Moreover, we illustrate the benefits of combining memory and storage window allocations, as a mechanism for seamlessly supporting hybrid memory hierarchies.

The contributions of this work are the following:
    
\begin{itemize}
\item We design and implement MPI \emph{storage} windows to map MPI windows into storage devices. We provide a reference, open-source implementation atop the MPI profiling interface, and consider how the approach could be integrated inside MPICH.
\item We show that MPI storage windows introduce only a relatively small runtime overhead when compared to MPI memory windows and MPI I/O, in most cases. However, it provides a higher level of flexibility and seamless integration of the memory and storage programming interfaces.
\item We present how to use MPI storage windows for out-of-core execution, parallel I/O and fault-tolerance in reference applications, such as Distributed Hash Table, HACC I/O kernel, and MapReduce ``One-Sided''.
\item We illustrate how heterogeneous window allocations can provide performance advantages when applications benefit from \emph{combined} memory and storage allocations.
\end{itemize}

The paper is organized as follows. We provide an overview of MPI windows and present the design and implementation of MPI storage windows in \autoref{2_Methods}. The experimental setup and performance results of the Intel MPI RMA Benchmarks, a STREAM-inspired microbenchmark, Distributed Hash Table, HACC I/O kernel, and MapReduce mini-applications are presented in \autoref{3_Results}. We extend the discussion of the results and provide further insights in \autoref{4_Discussion}. Related work is described in \autoref{5_RelatedWork}. Lastly, \autoref{6_Conclusion} summarizes our conclusions and outlines future work.

\section{MPI Storage Windows}
\label{2_Methods}

The MPI ``windows'' concept was introduced in MPI-2 to support the one-sided communication model. With MPI windows, a process can access the address space of local or remote processes without explicit \texttt{send} plus \texttt{receive} communication. The term \emph{window} is used because only a limited part of the local memory is exposed to other MPI processes. This is similar to a window in a window pane~\cite{gropp2014using}. The memory space that is not explicitly exposed through the MPI window still remains private, making the model safe against programming errors (e.g., buffer overflow).

The basic operations defined by the MPI standard to access and update an MPI window are \texttt{put} and \texttt{get}. These operations can be used in local or remote MPI processes. MPI-2 also introduced accumulate functions and synchronization operations on the exposed window. These synchronization operations are important to ensure the data availability after \texttt{put} and \texttt{get} operations. They also help to avoid race conditions on the window. MPI-3, on the other hand, extended the one-sided communication with functionality that supports the \mbox{\emph{passive}} target synchronization, consolidating the one-sided communication model by allowing decoupled interaction among the processes~\cite{gerstenberger2014enabling}. The new revision of the standard additionally defined atomic operations, such as Compare-And-Swap (CAS). It also extended the concept of MPI window by introducing MPI \emph{dynamic} windows and MPI \emph{shared} memory windows~\cite{gropp2014using, hoefler2013mpi}. The shared memory windows support direct \texttt{load} / \texttt{store} operations for intra-node communication. This represents an alternative to shared memory programming interfaces, such as OpenMP.

\begin{figure*}
    \centering
    \begin{subfigure}[t]{0.47921\textwidth}
        \centering
        \includegraphics[width=0.81\columnwidth]{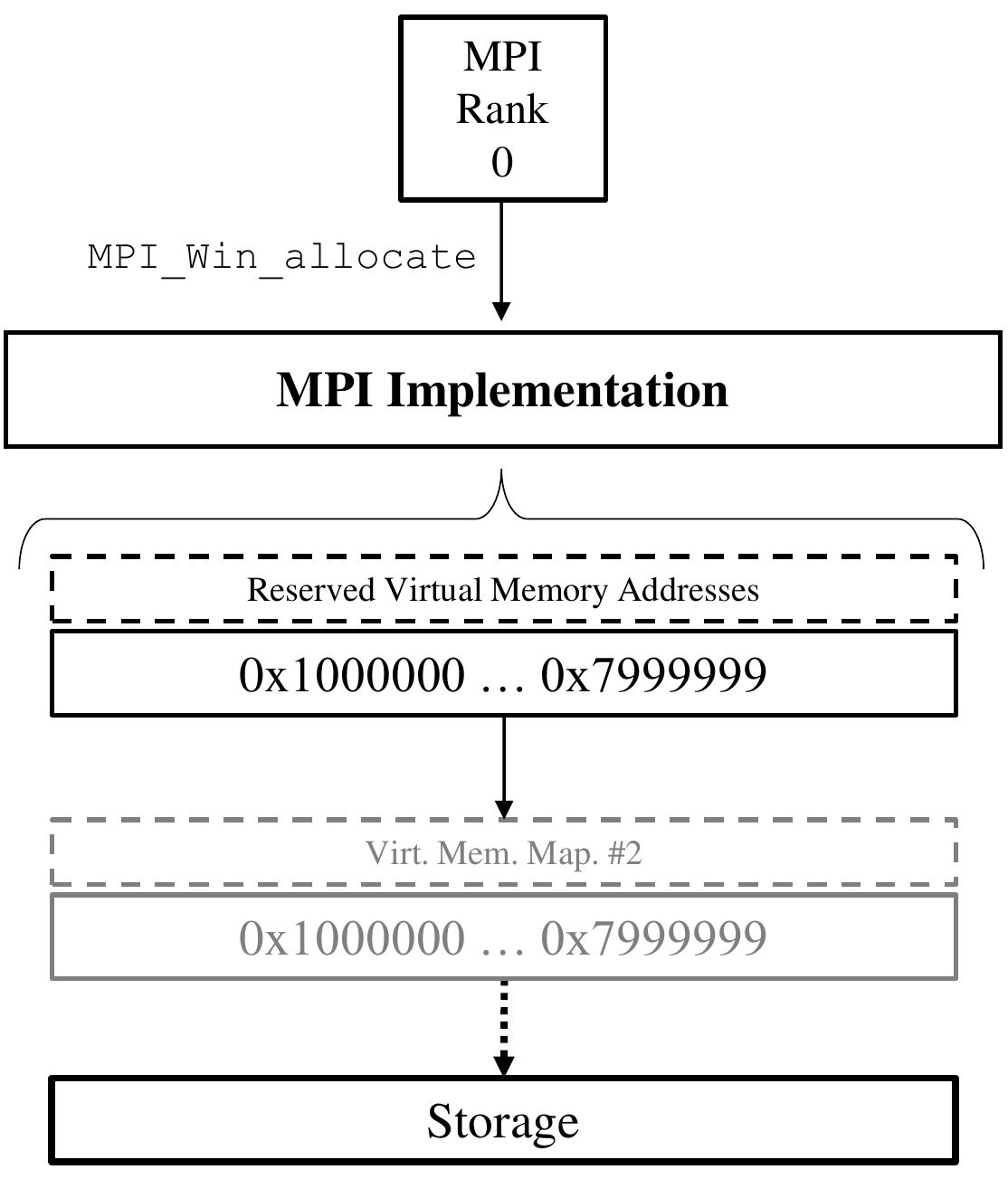}
        \caption{Storage window allocation} \label{fig:winalloc_storage}
    \end{subfigure}
    \hfill
    \begin{subfigure}[t]{0.47921\textwidth}
        \centering
        \includegraphics[width=0.81\columnwidth]{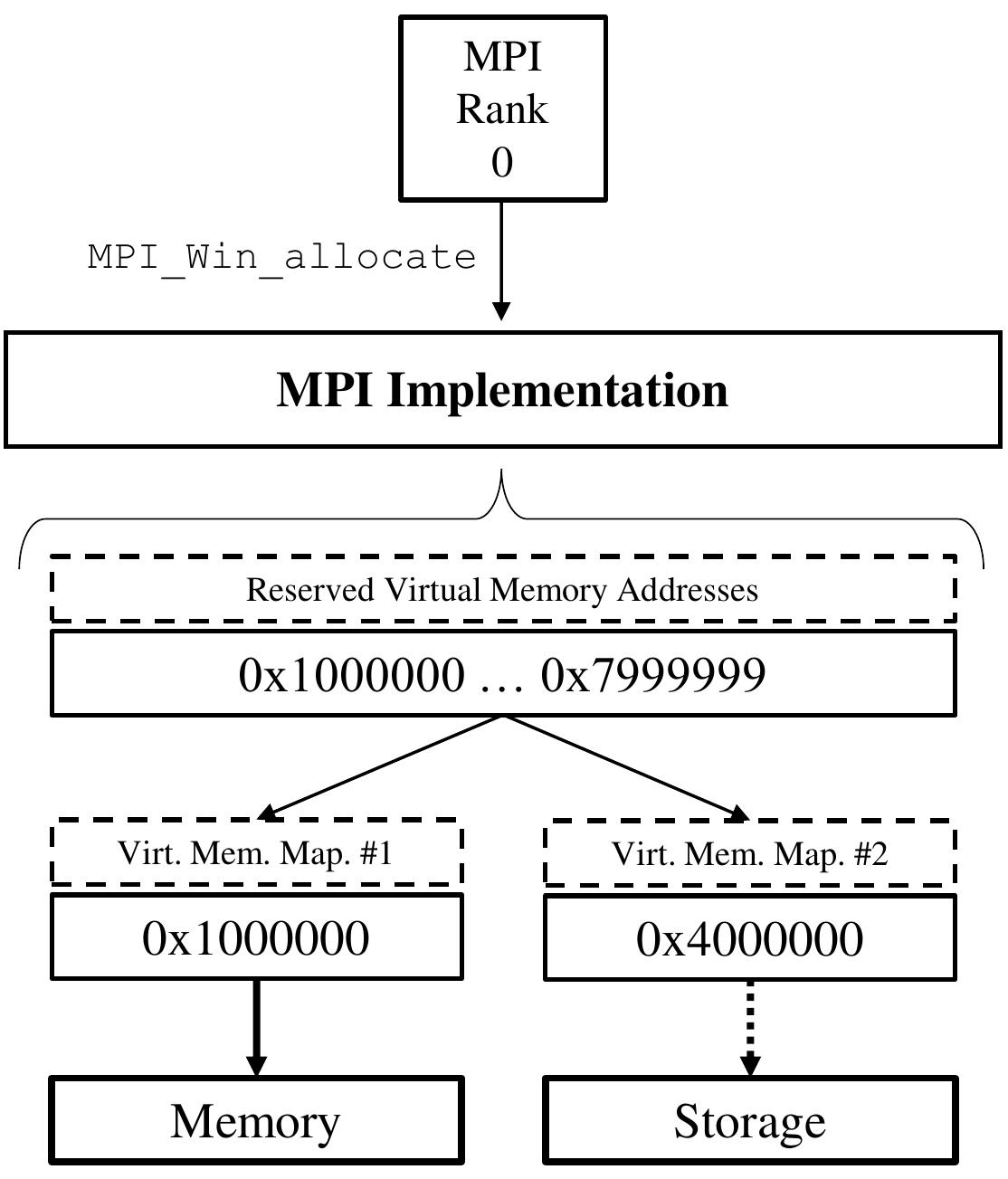}
        \caption{Combined window allocation} \label{fig:winalloc_combined}
    \end{subfigure}
    \caption{(\subref*{fig:winalloc_storage}) Storage window allocations are defined by reserving a range of virtual addresses and establishing a map to storage. (\subref*{fig:winalloc_combined}) Combined window allocations are defined by dividing the reserved range of virtual addresses, and then mapping each subrange individually. Thus, applications are provided with a single address space that contains both allocation types.\label{fig:winalloc_storage_combined}}
\end{figure*}

In this regard, extending the MPI window concept to storage requires no change to the MPI standard. The reason is that the standard does not restrict the type of allocation that an MPI window should be pinned to. Therefore, MPI windows can be easily allocated on storage if proper \emph{performance hints} are given via the MPI \emph{Info Object}. The performance hints are tuples of key / value pairs that provide an MPI implementation with information about the underlying hardware. For instance, certain hints can improve the performance of collective operations by providing network topology. Also, MPI I/O operations can be optimized if the I/O characteristics of an application are provided through hints. Thus, performance hints can determine the location of the mapping to storage and define other hardware-specific settings, such as the file striping size of Lustre. In addition, applications might configure allocations to combine memory and storage. This would create a unified virtual address space, where part of the window allocation is referring to memory and the rest to storage (\autoref{fig:winalloc_storage_combined}).

Support for MPI windows on storage can be ideally provided at MPI implementation level. The feature can be easily integrated at library-level as well. Understanding the type of allocation is possible through the \emph{attribute caching} mechanism of MPI. This feature enables user-defined \emph{cached} information on MPI communicators, datatypes and windows. In this case, metadata about the allocation attached to the MPI window object can be stored. Hence, it is possible to differentiate between traditional in-memory allocations, storage-based, or combined allocations. The location of the mapping and its properties can be retrieved by querying the MPI window object.

In this section, we present the design and implementation details of MPI storage windows. Note that, from now on, the term MPI \emph{memory} window is used to refer to the traditional MPI window allocation in memory. The term MPI \emph{storage} window is used to refer to our proposed extension. Additionally, the term \emph{combined} window allocation is used to refer to heterogeneous allocations.

\subsection{Design and Implementation}

We design and implement MPI storage windows as a library\footnote{\url{https://github.com/sergiorg-kth/mpi-storagewin}} on top of MPI using the MPI \emph{profiling interface}~\cite{karrels1994performance}. We also integrate the approach inside the MPICH MPI implementation (CH3)~\cite{gropp1996high}. The library version allows us to quickly prototype the MPI storage window concept and to understand which features are required for supporting storage-based allocations in the future. The MPICH integration allows us to understand the complexity of defining this concept in a production-quality MPI implementation. Here, we mostly re-use the existing code developed for MPI windows and expand the window structure with certain attributes (e.g., new window flavor). Nonetheless, both implementations support the same functionality, consist of approximately 500 lines of code, and feature identical performance. In this section, we will provide details about the main concepts behind both implementations.

We define seven different performance hints to enable and configure MPI storage windows. If the specific MPI implementation does not support storage allocations, the performance hints are simply ignored. These are the new hints introduced:

\begin{itemize}
    \item {\ttfamily alloc\_type}. If set to ``{\ttfamily storage}'', it enables the MPI window allocation on storage. Otherwise, the window will be allocated in memory (default).
    \item {\ttfamily storage\_alloc\_filename}. Defines the path and the name of the target file. A block device can also be provided, allowing us to support different storage technologies. In addition, shared files are allowed if the same target is defined among all the processes of the communicator.
    \item {\ttfamily storage\_alloc\_offset}. If the target file exists, the offset identifies the MPI storage window starting point. This offset is also valid when targeting block devices directly.
    \item {\ttfamily storage\_alloc\_factor}. Enables combined window allocations, where a single virtual address space contains both memory and storage. A value of ``{\ttfamily 0.5}'' would associate half of the addresses into memory, and half into storage. Using ``{\ttfamily auto}'' would set the correct allocation factor if the requested window size exceeds the main memory capacity.
    \item {\ttfamily storage\_alloc\_order}. Defines the order of the allocation when using the combined window allocations. A value of ``{\ttfamily memory\_first}'' sets the first part of the address space into memory, and the rest into storage (default).
    \item {\ttfamily storage\_alloc\_unlink}. If set to ``{\ttfamily true}'', it removes the associated file during the deallocation of an MPI storage window (i.e., useful for writing temporary files). This hint has no effect when targeting block devices.
    \item {\ttfamily storage\_alloc\_discard}. If set to ``{\ttfamily true}'', avoids to synchronize to storage the recent changes during the deallocation of the MPI storage window.
\end{itemize}

\begin{figure*}
    \centering
    \includegraphics[width=1.81\columnwidth]{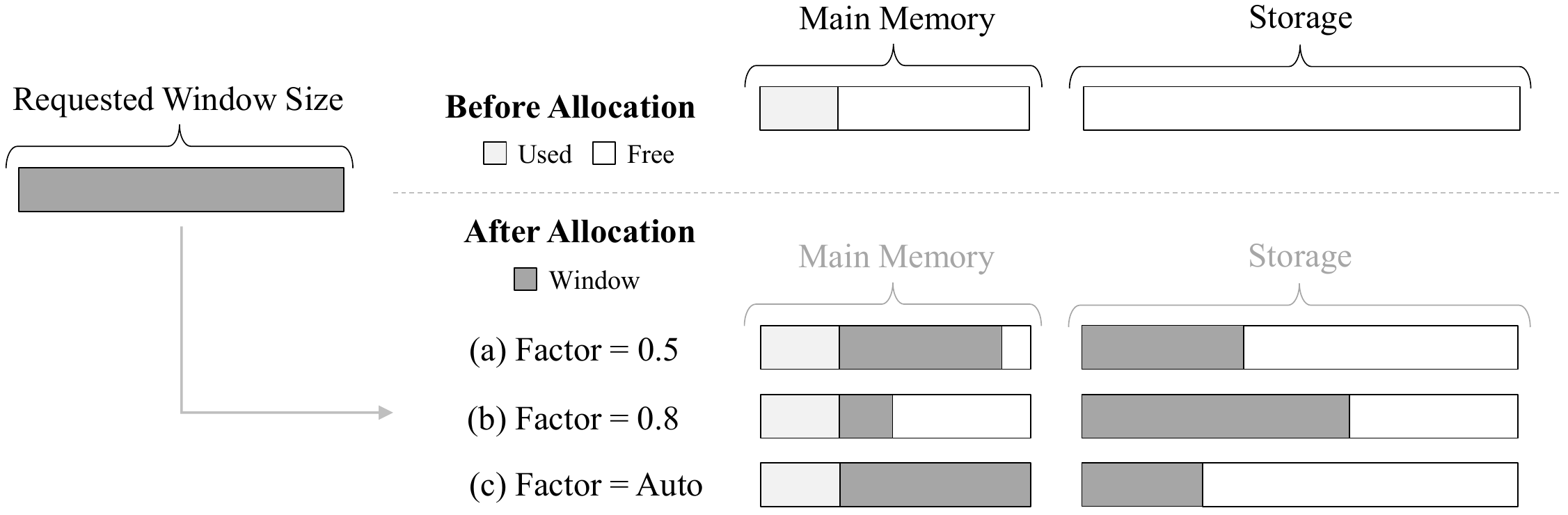}
    \vspace{0.0921cm}
    \caption{Combined window allocations are configured through a ``factor'' hint. The example illustrates allocating a certain window size using a factor of \texttt{0.5} (a), \texttt{0.8} (b), or \texttt{auto} (c). In this latter case, the aim is to let the MPI implementation decide the optimal factor.\label{fig:winalloc_storage_combined_factor}}
\end{figure*}

Applications that use MPI one-sided communication can continue to allocate windows in memory by avoiding to provide the {\ttfamily alloc\_type} hint, or by setting this hint with a value of ``{\ttfamily memory}''. To enable MPI storage windows, the {\ttfamily alloc\_type} hint has to be set to ``{\ttfamily storage}''. Applications are then expected to provide, at least, the {\ttfamily storage\_alloc\_filename} hint, which is required to specify the path where the window is set to be mapped (e.g., a file). The rest of the described hints are optional and will strictly depend on the particular use-case where MPI storage windows is integrated. 
For instance, using the {\ttfamily storage\_alloc\_factor}, part of the virtual memory address space can be divided into a traditional memory allocation plus a storage allocation, while still maintaining a unified virtual address space. The {\ttfamily storage\_alloc\_order} hint defines the order of the mapped addresses, that can correspond to memory first and then storage, or vice versa. Applications can additionally opt to define a factor value of ``{\ttfamily auto}'' for out-of-core execution using MPI storage windows. In such case, when the requested allocation exceeds the main memory capacity, the factor will be adapted to map the part that exceeds the main memory into storage. Otherwise, the window allocation remains in memory by default. \autoref{fig:winalloc_storage_combined_factor} illustrates the differences using a fixed factor of {\ttfamily 0.5}, {\ttfamily 0.8}, and finally {\ttfamily auto}.

We also integrate some of the reserved hints defined in the MPI I/O specification. These are mostly designed to optimize the data layout and access patterns on parallel file systems, such as Lustre. The hints supported are described below:

\begin{itemize}
    \item {\ttfamily access\_style}. Specifies the access pattern of the target file or block device used for the MPI storage window (e.g., ``{\ttfamily read\_mostly}'' when mainly read operations are required).
    \item {\ttfamily file\_perm}. Establishes the file permissions when creating a new file for the window. This hint has no effect when targeting existing files or block devices.
    \item {\ttfamily striping\_factor}. Defines the number of I/O devices that the MPI storage window should be striped across (e.g., number of OST devices on Lustre). This hint has no effect when targeting existing files or block devices.
    \item {\ttfamily striping\_unit}. Sets the striping unit to be used for the MPI storage window (e.g., stripe size of Lustre). This hint has no effect when targeting existing files or block devices.
\end{itemize}

\begin{figure*}
  \begin{center}
    \includegraphics[width=1.81\columnwidth]{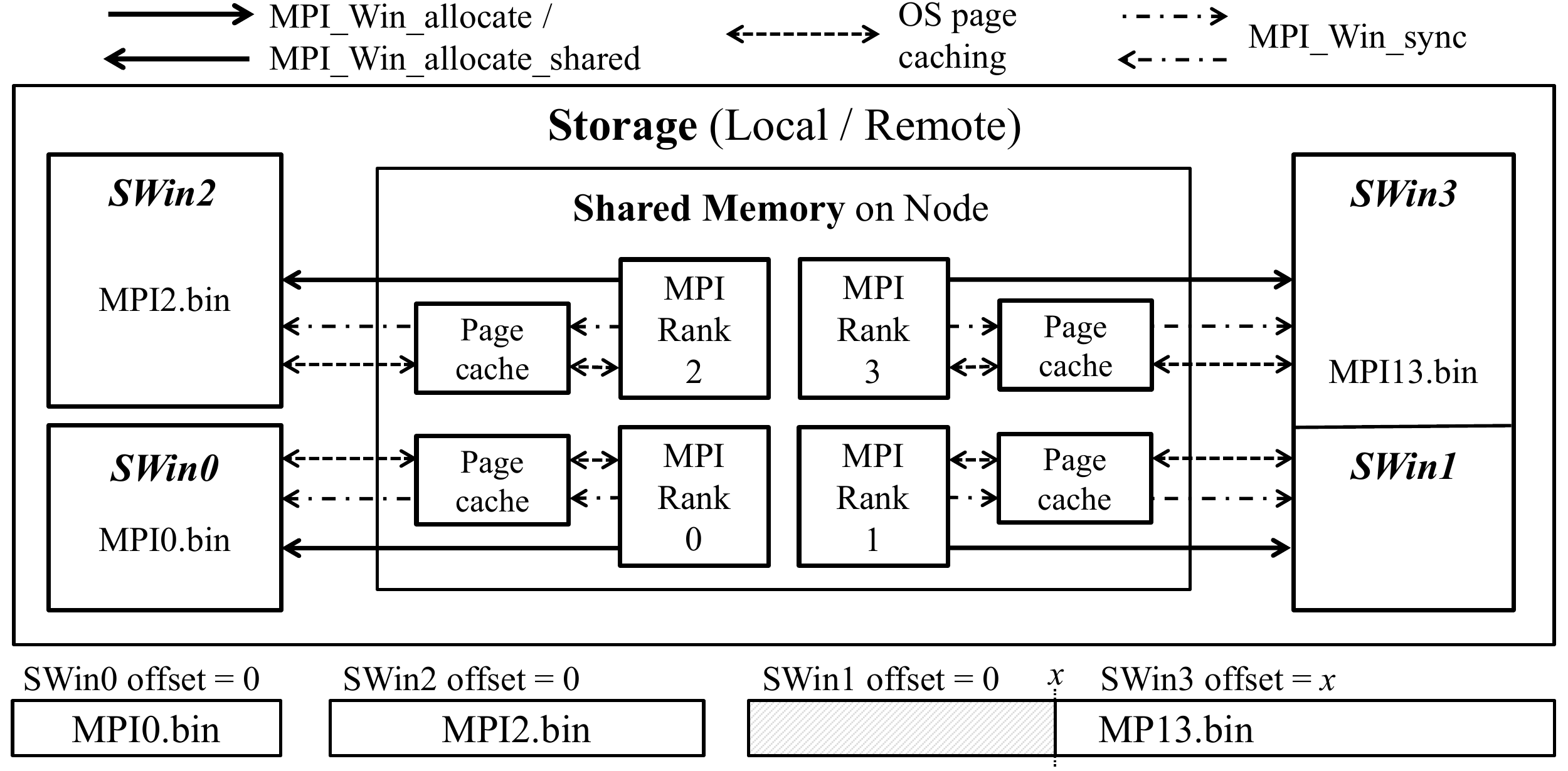}
    \vspace{0.0921cm}
    \caption{MPI storage windows can be created by mapping files into the MPI process address space through the memory-mapped I/O mechanism of the OS. The page cache optimizes the performance by maintaining part of the mapped space stored in memory. With \texttt{MPI\_Win\_sync}, the memory and storage copies of the window are guaranteed to be synchronized.}
    \label{diagramdesign}
  \end{center}
\end{figure*}

Our implementation of MPI storage windows is based on the use of memory-mapped file I/O~\cite{bovet2005understanding}. Target files or block devices from an MPI storage window are first opened, mapped into the virtual memory space of the MPI process, and then associated with the MPI window. A similar procedure is followed when creating combined window allocations. In this case, the allocation is separated in two steps, as previously illustrated in \autoref{fig:winalloc_storage_combined}. First, a range of virtual memory addresses that corresponds to the requested allocation size is reserved. This ensures that applications obtain a seamless virtual address space with the same base. Thereafter, this range is divided into individual mappings that point to memory and storage, respectively. The division and the order are determined through the performance hints. By default, the memory allocation appears first, unless otherwise specified.

For these purposes, five basic Unix system and I/O functions are required:
    
\begin{itemize}
    \item {\ttfamily mmap}. This system call is used to reserve the range of virtual addresses, as well as to map memory, files, and block devices into the virtual memory space of the MPI process. We use {\ttfamily MAP\_SHARED} to enable page sharing among different processes, {\ttfamily MAP\_NORESERVE} to avoid the use of swap space, and {\ttfamily MAP\_FIXED} to customize the range of virtual addresses for the final mapping. In case of memory allocations, we set {\ttfamily MAP\_ANONYMOUS} as well.
    \item {\ttfamily ftruncate}. When targeting files, this function is used to guarantee that the mapping has enough associated storage space. Otherwise, writing beyond the last mapped page would result in a segmentation fault.
    \item {\ttfamily msync}. This system call flushes all the dirty pages to storage from the page cache of the OS. We enforce the synchronous mode of this call, blocking the process until the data is guaranteed to be stored. 
    \item {\ttfamily munmap}. This system call releases memory allocations and removes the mapping of the file or block device from the page table of the process.
    \item {\ttfamily unlink}. This I/O function allows us to delete the mapped file from storage (e.g., during deallocation).
\end{itemize}

On the other hand, we extend the functionalities of several MPI routines to handle the allocation, deallocation, and synchronization of MPI storage windows. The interface of these routines remains unaltered and follow the original specification of the MPI standard. Thus, the programmer is only required to provide the Info Object with the described hints to enable MPI storage windows. The main routines extended are:

\begin{itemize}
    \item {\ttfamily MPI\_Win\_allocate}. This routine allocates an MPI storage window, taking as argument the MPI Info Object. The routine maps the physical file or block device into the virtual memory space of the MPI process, following the memory mapped file I/O mechanism. It also associates the mapping within the MPI window object as a cached attribute. The specified file in the performance hints is created if it does not exist, or resized if required. This routine performs a collective operation.
    \item {\ttfamily MPI\_Win\_allocate\_shared}. Equivalent to the previous routine, it defines an MPI shared window on storage. Hence, MPI processes have efficient access to the mapped storage of other processes within the same shared computing node. By default, the mapped addresses are consecutive, unless specified. This routine performs a collective operation.
    \item {\ttfamily MPI\_Win\_free}. This routine releases the mapping from the page table of the MPI process. If requested with the {\ttfamily storage\_alloc\_unlink} hint, it also deletes the mapped file. This routine performs a collective operation.
    \item {\ttfamily MPI\_Win\_sync}. This routine synchronizes the memory and storage copies of the MPI storage window\footnote{Even though MPI storage windows resembles the \emph{separate} memory model of MPI windows~\cite{gropp2014using}, in this case, local and remote operations only affect the memory mapped region.}. The window synchronization enforces the OS to write any dirty pages to storage. This routine may return immediately if the pages are already synchronized with storage by the OS (i.e., a selective synchronization is frequently performed).
    \item {\ttfamily MPI\_Win\_attach} / {\ttfamily MPI\_Win\_detach}. These two routines allow us to support MPI dynamic windows on storage. The routines can attach / detach a storage mapping from a given MPI dynamic window. The mapping to storage can be pre-established by providing the hints to {\ttfamily MPI\_Alloc\_mem}.
\end{itemize}

\autoref{diagramdesign} summarizes the use of these routines with four MPI storage windows. Three files are opened and mapped into the virtual memory space of the four MPI process. After the mapping is established, the OS automatically moves data from memory (\emph{page cache}) to storage, and vice-versa. In the example, two MPI storage windows share a file. An offset $x$ is provided as performance hint during the allocation of the window. The {\ttfamily MPI\_Win\_sync} ensures that the window copy on memory, within the page cache of the OS, is synchronized with the mapped files on storage.

\subsubsection{Data Consistency with MPI Storage Windows}

By using memory-mapped I/O, MPI storage windows implicitly integrate demand paging in memory through the \emph{page cache} of the OS. The page cache temporary holds frequently accessed pages mapped to storage. Hence, read operations trigger data accesses to storage only if the data is not available inside the page cache. Write operations can enforce a direct synchronization to storage, or aggregate the operations to increase the performance. 
The {vm.*} settings\footnote{\url{https://www.kernel.org/doc/Documentation/sysctl/vm.txt}} determine the interval and retention period of the pages stored within the page cache. The amount of active dirty pages in memory is specified through the \texttt{vm.dirty\_ratio} setting. A lower ratio will guarantee data consistency with storage at any time. A higher ratio will absorb bursts of sequential or consecutive writes over a certain memory region, improving the performance. The OS will always continue to flush all the dirty pages in the background. The frequency of these flushes can be configured with the \texttt{vm.dirty\_writeback\_centisecs} setting.

\begin{lstlisting}[float=*,caption={Allocation of MPI storage windows and writing data to remote processes.}, label={exampleCode}]
...
// Define the MPI_Info object to enable storage allocations 
MPI_Info_create(&info); 
MPI_Info_set(info, "alloc_type", "storage");                   
MPI_Info_set(info, "storage_alloc_filename", "/path/tofile"); 
MPI_Info_set(info, "storage_alloc_offset", "0");
MPI_Info_set(info, "storage_alloc_unlink", "false"); 
   
// Allocate storage window with space for num_procs integers
MPI_Win_allocate(num_procs * sizeof(int), sizeof(int), info,  
                 MPI_COMM_WORLD, (void**)&baseptr, &win);

if (IS_EVEN_NUM(rank)) {
  for(int drank = 1; drank < num_procs; drank += 2) {
    // Put our own rank plus some number to the dest. process 
    int k = rank + timestamp;
    MPI_Win_lock(MPI_LOCK_SHARED, drank, 0, win);
    MPI_Put(&k, 1, MPI_INT, drank, offset, 1, MPI_INT, win);
    MPI_Win_unlock(drank, win);
  }
}
...
\end{lstlisting}

\begin{lstlisting}[float=*,caption={Heterogeneous allocation that combines 50\% memory and 50\% storage.}, label={exampleCode_factor}]
...
// Define the MPI_Info object to enable combined allocations 
MPI_Info_create(&info); 
MPI_Info_set(info, "alloc_type", "storage");                   
...
MPI_Info_set(info, "storage_alloc_factor", "0.5");
   
// Allocate the window with half the space in memory
MPI_Win_allocate(num_procs * sizeof(int), sizeof(int), info,  
                 MPI_COMM_WORLD, (void**)&baseptr, &win);
...
\end{lstlisting}

\begin{lstlisting}[float=*,caption={Allocation of MPI dynamic windows on storage.}, label={exampleCode_dynamic}]
...
// Define the MPI_Info object to enable storage allocations 
MPI_Info_create(&info); 
MPI_Info_set(info, "alloc_type", "storage");
...
// Allocate space for num_procs integers on storage
MPI_Alloc_mem(num_procs * sizeof(int), info, (void**)&baseptr);

// Create the dynamic window and attach the storage allocation
MPI_Win_create_dynamic(MPI_INFO_NULL, MPI_COMM_WORLD, &win);
MPI_Win_attach(win, baseptr, num_procs * sizeof(int));
...
\end{lstlisting}

The flexibility of memory mapped I/O, however, also introduces several challenges for data consistency inside MPI storage windows. First and foremost, local or remote operations are \emph{only} guaranteed to affect the memory copy of the window inside the page cache. The semantics of the MPI one-sided communication operations, such as \texttt{MPI\_Win\_lock} / \texttt{MPI\_Win\_unlock} or \texttt{MPI\_Win\_flush}, will only ensure the completion of the local or remote operations inside the memory of the target process (i.e., the storage status is undefined at that point). As a consequence, when consistency of the data within the storage layer has to be preserved, applications are required to use {\ttfamily MPI\_Win\_sync} on the window to enforce a synchronization of the modified content inside the page cache of the OS. This operation blocks the MPI process until the data is ensured to be flushed from memory to the storage device. Even though our current implementation only supports a local synchronization of the window (i.e., intra-node), we consider that the semantics of this operation should trigger a storage synchronization on remote processes as well. Therefore, write operations (e.g., \texttt{MPI\_Put}) accompanied with a subsequent {\ttfamily MPI\_Win\_sync} will guarantee data consistency on the storage layer of the remote process. Read operations (e.g., \texttt{MPI\_Get}), on the other hand, are not affected and will trigger data accesses to storage through the page fault mechanism of the OS.

The second challenge is to prevent remote data accesses during a window synchronization to storage. In this regard, the MPI standard already contains the definition of an \emph{exclusive} lock inside the passive target synchronization~\cite{MPI3standard2015}. By default, locks are used to protect accesses to the target window, and to protect local \texttt{load} / \texttt{store} accesses to a locked local window. Accesses that are protected by an exclusive lock (i.e., \texttt{MPI\_LOCK\_EXCLUSIVE}) will not be concurrent with other accesses to the same window that are lock protected. Thus, guaranteeing that no interference exists during the synchronization of the data.

We must note that, in the future, MPI implementations might opt to use a different approach to implement MPI storage windows, such as MPI I/O. This would allow to have full data consistency control at MPI implementation-level. For instance, data consistency might only be possible through {\ttfamily MPI\_Win\_sync}, without involving the OS.

\subsection{Using MPI Storage Windows}

MPI applications that use the MPI one-sided communication model can immediately use MPI storage windows. \autoref{exampleCode} shows a code example that allocates an MPI storage window using some of the described performance hints. The example demonstrates how different MPI processes can write information to an MPI storage window of other processes with a simple \texttt{put} operation. This operation can also be used for local MPI storage windows. The code first creates an MPI Info Object and sets the performance hints to enable storage allocations. Then, it allocates the MPI storage window and instruct each even-rank MPI process to write to the correspondent MPI storage window of odd-rank MPI processes. The {\ttfamily MPI\_Win\_lock} and {\ttfamily MPI\_Win\_unlock} start and end the passive epoch to the MPI window on the target rank.

The \texttt{baseptr} pointer returned by the window allocation call can be used for local \texttt{load} / \texttt{store} operations. In MPI shared windows on storage, processes can \texttt{load} / \texttt{store} data from each other by pointer dereferencing.

On the other hand, \autoref{exampleCode_factor} illustrates how to create a window that combines memory and storage. We can use the ``factor'' hint to specify the data distribution inside the allocation. In the example, half of the space is allocated in memory.

The MPI standard additionally defines MPI dynamic windows, that allow applications to dynamically attach memory allocations after the window is created. \autoref{exampleCode_dynamic} illustrates how MPI dynamic windows on storage can be defined. The performance hints are provided to \texttt{MPI\_Alloc\_mem} instead.


\section{Experimental Results}
\label{3_Results}

In this section, we illustrate the performance of MPI storage windows using two different testbeds. The first testbed is a single computing node with local storage (hard disk and SSD), which allows us to demonstrate the implications of MPI storage windows on upcoming clusters with local persistency support. The second testbed is a supercomputer at KTH Royal Institute of Technology, with storage provided by a Lustre parallel file system. This allows us to understand how MPI storage windows can be integrated into current HPC clusters with network-based storage support. The specifications are described below:

\begin{itemize}
\item {\bf Blackdog} is a workstation with an eight-core Xeon E5-2609v2 processor running at 2.5GHz. The workstation is equipped with a total of 72GB DRAM. The storage consists of two 4TB HDD (WDC WD4000F9YZ / non-RAID) and a 250GB SSD (Samsung 850 EVO). The OS is Ubuntu Server 16.04 with Kernel 4.4.0-62-generic. The applications are compiled with {\ttfamily gcc} v5.4.0 and MPICH v3.2.
\item {\bf Tegner} is a supercomputer with 46 compute nodes that are equipped with Haswell E5-2690v3 processor running at 2.6GHz. Each node has two sockets with 12 cores and a total of 512GB DRAM. The storage employs a Lustre parallel file system (client v2.5.2) with 165 OST servers. No local storage is provided per node. The OS is CentOS v7.3.1611 with Kernel 3.10.0-514.6.1.el7.x86\_64. The applications are compiled with {\ttfamily gcc} v6.2.0 and Intel MPI v5.1.3.
\end{itemize}

Using these two testbeds, we first verify that MPI storage windows does not incur in additional overheads when performing remote memory operations on the page cache compared to MPI memory windows. We then estimate the throughput of large memory operations using a custom microbenchmark that enforces synchronization to storage. We also present example applications that can take advantage of MPI storage windows, and compare the performance with MPI I/O. Lastly, we provide insights into novel techniques that could take advantage of the approach, such as transparent checkpointing. After this section, we continue and extend the discussion on the obtained results.

Note that all the figures reflect the standard deviation of the samples as error bars. We use the PMPI-based implementation for our experiments in both testbeds, mainly due to deployment reasons on Tegner\footnote{We verify on Blackdog that no performance differences exist between the MPICH implementation of MPI storage windows and the PMPI implementation.}. We increase the \texttt{vm.dirty\_ratio} setting of the OS to 80\% on Blackdog to allow for a higher amount of dirty pages in memory. In addition, we set the default Lustre settings on Tegner, assigning one OST server per MPI process and a stripping size of 1MB. The swap partition is disabled in both testbeds. Lastly, the evaluations on Tegner are conducted on different days and timeframes, to account for the interferences produced by other users on the cluster.

\subsection{Intel IMB-RMA Benchmarks}

For our first evaluation, we verify whether MPI storage windows can incur in performance overheads using MPI one-sided operations. This might be the case where special memory and techniques are required for data consistency on RDMA~\cite{mitchell2013using}. The goal is to ensure that no subtle performance differences exist for small data transfers using the page cache in comparison with MPI windows allocated directly in memory (i.e., without storage synchronization). For this purpose, we use the Intel MPI Benchmarks (IMB), an open-source project\footnote{\url{https://github.com/intel/mpi-benchmarks}} that performs a set of performance measurements that fully characterizes the efficiency of MPI implementations. In particular, we use the IMB-RMA subset of benchmarks to measure the throughput of the one-sided operations of the MPI-3 standard.

The IMB-RMA is divided into three sets: ``Single Transfer'' (one process accesses the memory of another process), ``Multiple Transfer'' (one process accesses the memory of several other processes), and ``Parallel Transfer'' (multiple processes transfer data in parallel). We select a subset of these benchmarks that better characterizes the performance of our implementation. We introduce the necessary performance hints to enable MPI storage windows, maintaining the original code unaltered. In our tests, we set the \emph{standard} mode to allow a single group of processes in parallel. We also configure the benchmarks in \emph{non-aggregate} mode to make sure that only one RMA operation is performed on each passive target epoch. The transfer sizes per test vary in powers of two, from 256KB up to 4MB. In addition, we disable the \emph{iteration policy}, fixing in 1000 iterations per test, regardless of transfer size.

\input{3_Results_1_IMB_fig1}

\input{3_Results_1_IMB_fig2}

Using the ``Single Transfer'' set of the IMB-RMA benchmarks, we determine that the use of memory-mapped I/O on MPI storage windows only incurs in a small performance overhead when performing small unidirectional data transfers, and a negligible overhead on small bidirectional data transfers and atomic RMA operations. \autoref{fig:imb_single} shows the throughput achieved using MPI memory windows and MPI storage windows on Tegner. The tests use 2 MPI processes split into 2 separate nodes of the cluster. Each storage window is mapped as an independent file per process in Lustre. Figures~\ref*{fig:imb_single}\subref*{fig:imb_unidirput} and \ref*{fig:imb_single}\subref*{fig:imb_unidirget} indicate that subtle performance differences on unidirectional \texttt{put} and \texttt{get} operations occur based on the transfer size selected. For instance, using 1MB data transfer for unidirectional \texttt{put} operations, we observe an average throughput of 6.87GB/s for MPI memory windows, and 6.95GB/s for MPI storage windows (i.e., approximately 1\% difference). Figures~\ref*{fig:imb_single}\subref*{fig:imb_bidirput} and \ref*{fig:imb_single}\subref*{fig:imb_bidirget} illustrate similar results when performing bidirectional \texttt{put} and \texttt{get} operations. For the bidirectional \texttt{put} test, we observe a peak of 5.24GB/s on average for MPI memory windows, and a peak of 5.20GB/s for MPI storage windows (i.e., 0.06\% difference). For the bidirectional \texttt{get} test, the peak transfer rate is 5.14GB/s for MPI memory windows and 5.20GB/s for MPI storage windows (i.e., 1\% difference). Finally, atomic operations, such as Accumulate (\subref*{fig:imb_accumulate}), Get-Accumulate (\subref*{fig:imb_getaccumulate}), Fech-and-Op (\subref*{fig:imb_fetchandop}) and Compare-and-Swap (\subref*{fig:imb_cas}), show no relevant performance differences on each test. In both cases, the average throughput is reduced (e.g., 3.09GB/s for Get-Accumulate).

On the other hand, using the ``Multiple Transfer'' and ``Parallel Transfer'' benchmarks of IMB-RMA, we determine that increasing the process count does not necessarily affect the performance for small data transfers when using MPI storage windows. \autoref{fig:imb_multiple_parallel} shows the throughput / execution time achieved with MPI memory windows and MPI storage windows on Tegner. The tests use 128 MPI processes split into 6 separate nodes of the cluster. Each storage window is again mapped into an independent file per process in Lustre. \autoref*{fig:imb_oneputall} indicates that MPI storage windows does not incur in any penalty while performing \texttt{put} operations over multiple processes, with a peak throughput of 5.89GB/s for MPI memory windows and 6.21GB/s for MPI storage windows. In the case of \texttt{get} operations, \autoref*{fig:imb_onegetall} shows a peak throughput of 1.49GB/s for MPI memory windows, and a peak of 1.44GB/s for MPI storage windows. The execution times of the Exchange-Put (\subref*{fig:imb_exchangeput}) and Exchange-Get (\subref*{fig:imb_exchangeput}) benchmarks confirm equivalent execution times on both implementations.

\subsection{mSTREAM Microbenchmark}

In order to understand the performance considerations of using MPI storage windows with storage synchronization enforcement, we define a custom microbenchmark inspired by STREAM~\cite{mccalpin1995survey}, called mSTREAM\footnote{\url{https://github.com/sergiorg-kth/mpi-storagewin/tree/master/benchmark}}. The purpose of this microbenchmark is to measure the access throughput by performing large memory operations over the window allocation. Thus, we aim to stress both the memory and storage subsystems to represent the worst performance scenario for memory-mapped IO (i.e., when no computations are conducted). The results are then compared with MPI memory windows.

\begin{figure}[t]
    \centering
    \begin{subfigure}[t]{0.47921\textwidth}
        \centering
        \hspace{-0.21cm}
        \begin{tikzpicture}
            \begin{axis}[
                xlabel=Kernel Type,
                ylabel=Throughput (MB/s),
                symbolic x coords={SEQ, PAD, RND, MIX},
                xtick=data,
                ytick={0,3000,6000,9000,12000},
                ymin=0,
                ymax=12000,
                scaled y ticks = false,
                y tick label style={/pgf/number format/fixed, /pgf/number format/1000 sep = },
                ylabel style={at={(-0.021,0.5621)}, style={font=\small}},
                xlabel style={at={(0.5,-0.05)}, style={font=\small}},
                enlarge x limits=0.21,
                grid=major,
                legend style={legend columns=2,at={(1.0,1.581)},anchor=north east},
                legend style={/tikz/every even column/.append style={column sep=4.21}},
                legend cell align=right,
                legend plot pos=right,
                legend style={draw=none, fill=white, inner xsep=0, inner ysep=0},
                ybar=0pt,
                bar width=5.81pt,
                width=0.8921\textwidth,
                height=3.4921cm,
                area legend
            ]
            \addplot [fill=black!21, postaction={pattern=north east lines}] plot[error bars/.cd, y dir=both, y explicit] table [y error minus=min, y error plus=max] {
                x       y               min             max
                SEQ     5668.8502318    56.7931031726   56.7931031726
                PAD     5744.1128398    34.0567291212   34.0567291212
                RND     5422.7559112    50.3021931744   50.3021931744
                MIX     5492.4284452    46.6531763237   46.6531763237
            };\addlegendentry{Memory}
            \addplot [fill=black!9, postaction={pattern=crosshatch dots}] plot[error bars/.cd, y dir=both, y explicit] table [y error minus=min, y error plus=max] {
                x       y               min             max
                SEQ     1849.1512776    394.3237849464  394.3237849464
                PAD     2785.2120876    67.0422633199   67.0422633199
                RND     2618.1968092    70.9077947459   70.9077947459
                MIX     2466.3405122    54.6485398174   54.6485398174
            };\addlegendentry{Storage (SSD)}
            
            \addlegendimage{empty legend}
            \addlegendentry{}
            
            \addplot [fill=black!5, postaction={pattern=dots}] plot[error bars/.cd, y dir=both, y explicit] table [y error minus=min, y error plus=max] {
                x       y               min             max
                SEQ     1706.7487198    100.5359642445  100.5359642445
                PAD     2524.2443946    101.4174940933  101.4174940933
                RND     2072.5786962    35.1454837271   35.1454837271
                MIX     2047.3508808    68.8162199011   68.8162199011
            };\addlegendentry{Storage (HDD)}
            \end{axis}
        \end{tikzpicture}
        \vspace{-0.21cm}
        \caption{\textbf{Blackdog Results} / 1 MPI Process} \label{fig:benchmark_bw_blackdog}
    \end{subfigure}
    
    \vspace{0.3921cm}
    
    \begin{subfigure}[t]{0.47921\textwidth}
        \centering
        \hspace{-0.21cm}
        \begin{tikzpicture}
            \begin{axis}[
                xlabel=Kernel Type,
                ylabel=Throughput (MB/s),
                symbolic x coords={SEQ, PAD, RND, MIX},
                xtick=data,
                ytick={0,3000,6000,9000,12000},
                ymin=0,
                ymax=12000,
                scaled y ticks = false,
                y tick label style={/pgf/number format/fixed, /pgf/number format/1000 sep = },
                ylabel style={at={(-0.021,0.5621)}, style={font=\small}},
                xlabel style={at={(0.5,-0.05)}, style={font=\small}},
                enlarge x limits=0.21,
                grid=major,
                legend style={legend columns=1,at={(1.0,1.581)},anchor=north east},
                legend style={/tikz/every even column/.append style={column sep=4.21}},
                legend cell align=right,
                legend plot pos=right,
                legend style={draw=none, fill=white, inner xsep=0, inner ysep=0},
                ybar=0pt,
                bar width=5.81pt,
                width=0.8921\textwidth,
                height=3.4921cm,
                area legend
            ]
            \addplot [fill=black!54] plot[error bars/.cd, y dir=both, y explicit] table [y error minus=min, y error plus=max] {
                x       y               min             max
                SEQ     10761.9699346   83.7909841489   83.7909841489
                PAD     11088.564117    101.60887904    101.60887904
                RND     10148.4290766   123.515528116   123.515528116
                MIX     10263.3367934   72.7346225049   72.7346225049
            };\addlegendentry{Memory}
            \addplot [fill=black!21] plot[error bars/.cd, y dir=both, y explicit] table [y error minus=min, y error plus=max] {
                x       y               min             max
                SEQ     846.2173242     45.0242460873   45.0242460873
                PAD     868.5976124     48.4561556867   48.4561556867
                RND     958.138817      38.7252254353   38.7252254353
                MIX     947.0190864     48.8732941296   48.8732941296
            };\addlegendentry{Storage (Lustre)}
            \end{axis}
        \end{tikzpicture}
        \vspace{-0.21cm}
        \caption{\textbf{Tegner Results} / 1 MPI Process} \label{fig:benchmark_bw_tegner}
    \end{subfigure}
    \caption{mSTREAM microbenchmark performance using MPI memory windows and MPI storage windows running on Blackdog and Tegner.\label{fig:benchmark_blackdog_tegner}}
\end{figure}
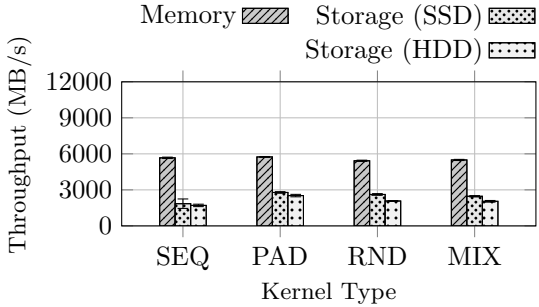
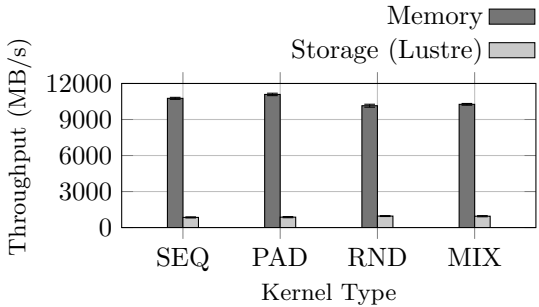

For this purpose, we define four different kernels that represent different type of memory accesses: sequential memory accesses (\texttt{SEQ}), sequential memory accesses with padding (\texttt{PAD}), pseudo-random memory accesses (\texttt{RND}), and mixed memory accesses (\texttt{MIX}), that combines the previous two. Each kernel performs accesses using large, configurable length segments. The benchmark alternates \texttt{read} / \texttt{write} operations. Data reuse is allowed in all the kernels except for \texttt{SEQ}. Several iterations are executed per kernel, and the number of memory operations per iteration strictly depends on the window allocation size. Hence, considering $M$ the size of the allocation and $S$ the segment size, the number of operations is determined by $M \div S$. The result is provided as average bandwidth $BW = \left(M \times I\right) \div \Delta{t}$, where $I$ is the number of iterations per kernel and $\Delta{t}$ is the elapsed execution time.

We configure the mSTREAM microbenchmark with a fixed window allocation size of 16GB and a segment size of 16MB per memory operation. This amounts to 1024 memory operations per iteration (512 \texttt{read} / 512 \texttt{write}). The number of iterations per kernel is set to 10 plus an initial ``cold'' iteration that does not count for the results. The amount of data transferred per test is 160GB (640GB in total). We instruct MPI to extend the window allocation to storage using the described performance hints. A single MPI process is used for each test to avoid potential interferences that could affect the results. In addition, we enforce a window synchronization point to storage before finishing the last iteration of each kernel.

\begin{figure}
    \centering
    \begin{subfigure}[t]{0.47921\textwidth}
        \centering
        \begin{tikzpicture}
            \begin{axis}[
                xlabel=Kernel Type,
                ylabel=Execution Time \%,
                symbolic x coords={SEQ, PAD, RND, MIX},
                xtick=data,
                ytick={0,0.2,0.4,0.6,0.8,1},
                ymin=0,
                ymax=1,
                scaled y ticks = false,
                y tick label style={/pgf/number format/fixed, /pgf/number format/1000 sep = },
                ylabel style={at={(0.081,0.4921)}, style={font=\small}},
                xlabel style={at={(0.5,-0.05)}, style={font=\small}},
                enlarge x limits=0.21,
                grid=major,
                legend style={legend columns=1,at={(1.0,1.581)},anchor=north east},
                legend style={/tikz/every even column/.append style={column sep=4.21}},
                legend cell align=right,
                legend plot pos=right,
                legend style={draw=none, fill=white, inner xsep=0, inner ysep=0},
                ybar=0pt,
                bar width=5.81pt,
                width=0.8921\textwidth,
                height=3.4921cm,
                area legend
            ]
            \addplot [fill=black!9, postaction={pattern=crosshatch dots}] plot[error bars/.cd, y dir=both, y explicit] table [y error minus=min, y error plus=max] {
                x       y               min             max
                SEQ     0.3270668603    0.0081394379    0.0081394379
                PAD     0.2497663105    0.0053551979    0.0053551979
                RND     0.3844126199    0.0090882782    0.0090882782
                MIX     0.3914824218    0.0023894532    0.0023894532
            };\addlegendentry{Storage (SSD)}
            \addplot [fill=black!5, postaction={pattern=dots}] plot[error bars/.cd, y dir=both, y explicit] table [y error minus=min, y error plus=max] {
                x       y               min             max
                SEQ     0.5056287109    0.00467038      0.00467038
                PAD     0.3575343262    0.0300389323    0.0300389323
                RND     0.4928121046    0.0061342153    0.0061342153
                MIX     0.5098654544    0.0163139607    0.0163139607
            };\addlegendentry{Storage (HDD)}
            \end{axis}
        \end{tikzpicture}
        \vspace{-0.21cm}
        \caption{\textbf{Avg. Flush Time} on Blackdog} \label{fig:benchmark_flush_blackdog}
    \end{subfigure}
    
    \vspace{0.3921cm}
    
    \begin{subfigure}[t]{0.47921\textwidth}
        \centering
        \begin{tikzpicture}
            \begin{axis}[
                xlabel=Operation,
                ylabel=Throughput (MB/s),
                symbolic x coords={Read, Write},
                xtick=data,
                ytick={0,2000,4000,6000,8000},
                ymin=0,
                ymax=8000,
                scaled y ticks = false,
                y tick label style={/pgf/number format/fixed, /pgf/number format/1000 sep = },
                ylabel style={at={(-0.021,0.4921)}, style={font=\small}},
                xlabel style={at={(0.5,-0.05)}, style={font=\small}},
                enlarge x limits=0.81,
                grid=major,
                legend style={legend columns=1,at={(1.0,1.364)},anchor=north east},
                legend style={/tikz/every even column/.append style={column sep=4.21}},
                legend cell align=right,
                legend plot pos=right,
                legend style={draw=none, fill=white, inner xsep=0, inner ysep=0},
                ybar=0pt,
                bar width=5.81pt,
                width=0.64\textwidth,
                height=3.4921cm,
                area legend
            ]
            \addplot [fill=black!21] plot[error bars/.cd, y dir=both, y explicit] table [y error minus=min, y error plus=max] {
                x       y               min             max
                Read    6986.2281294    322.2979302789  322.2979302789
                Write   419.2369286     29.9754495131   29.9754495131
            };\addlegendentry{Storage (Lustre)}
            \end{axis}
        \end{tikzpicture}
        \vspace{-0.21cm}
        \caption{\textbf{Avg. Throughput} on Tegner} \label{fig:benchmark_rw_tegner}
    \end{subfigure}
    \caption{(\subref*{fig:benchmark_flush_blackdog}) Normalized flushing time with MPI storage windows on Blackdog. (\subref*{fig:benchmark_rw_tegner}) Read / Write performance using the \texttt{SEQ} kernel with MPI storage windows on Tegner.\label{fig:benchmark_blackdog_flush_tegner_rw}}
\end{figure}
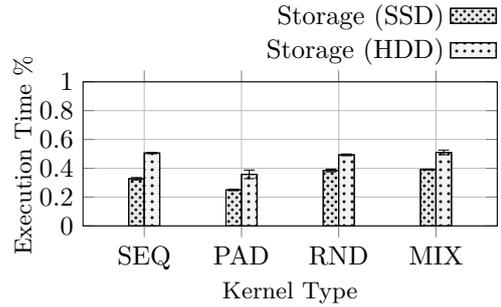
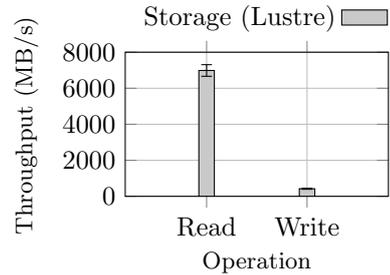

The performance of MPI storage windows is largely dominated by the need for flushing the data changes to storage when no other computations are conducted. \autoref{fig:benchmark_bw_blackdog} illustrates the throughput achieved by each kernel using MPI memory windows and MPI storage windows on Blackdog. The storage windows are mapped into the local storage of Blackdog, labelled ``SSD'' for the solid-state drive and ``HDD'' for the conventional hard disk. The performance results indicate that the degradation of storage window allocations is approximately 56.4\% on average when using the SSD. The maximum peak throughput observed is 2.72GB/s on the \texttt{PAD} kernel, in comparison with 5.61GB/s for MPI memory windows. However, \autoref{fig:benchmark_flush_blackdog} demonstrates that 33.8\% of the overall kernel execution time is spent on transferring the unflushed changes to storage. In the case of the conventional HDD, the flush time exceeds 50\% of the execution time in some of the kernels. Nonetheless, despite the bandwidth differences between the memory and storage subsystems, the page cache of the OS is effectively absorbing most of the I/O operations inside the cache. This fact is deducted by observing the peak theoretical bandwidth of the SSD used in our tests, expected to max out at 0.51GB/s (i.e., far below the rates measured).

On the other hand, mapping files to a Lustre parallel file system can constrain the performance. \autoref{fig:benchmark_bw_tegner} shows the results of the same experiment running on Tegner. In this case, MPI storage windows are mapped into a mounted Lustre parallel file system. Thus, I/O operations require network communication. We observe evident differences between MPI memory windows and MPI storage windows. The storage-based allocation degrades the throughput by over 90\% compared to the MPI window allocation in memory. In particular, the throughput of MPI memory windows reaches 10.6GB/s on average, while the peak throughput of MPI storage windows is approximately 1.0GB/s. The reason for this result is due to the lack of write cache on Lustre for memory-mapped I/O, that produces ``asymmetric'' performance for \texttt{read} and \texttt{write} operations on Tegner. We also confirm this effect by measuring the throughput of \texttt{read} and \texttt{write} with the \texttt{SEQ} kernel. We read from an MPI storage window to a memory-based, and from an MPI memory window to a storage-based, respectively. The result from this experiment is illustrated in \autoref{fig:benchmark_rw_tegner}.

\subsection{Distributed Data Structures}

Data analytics and machine learning applications are emerging on HPC~\cite{reed2015exascale, coates2013deep}. These applications pose a relatively large stress to the I/O subsystem, due to the large amounts of small I/O transactions that they produce. For our third performance evaluation, we use a Distributed Hash Table (DHT) implementation by Gerstenberger et al.~\cite{gerstenberger2014enabling}\footnote{\url{https://spcl.inf.ethz.ch/Research/Parallel_Programming/foMPI}} that mainly uses MPI one-sided operations. Hence, we intent to mimic data analytics applications that have random accesses to distributed data structures.

In this implementation, each MPI process handles a part of the DHT, named \emph{Local Volume} (LV). The processes also maintain an overflow heap to store elements in case of collisions. The LV and the overflow heap are allocated as MPI windows on each process, so that updates to the DHT are handled using solely MPI one-sided operations. In this way, each MPI process can \texttt{put} or \texttt{get} values, and also resolve conflicts asynchronously on any of the exposed LVs through Compare-and-Swap (CAS) atomic operations.

We introduce the necessary performance hints to map the content of the LV and heap to storage as MPI storage windows, using one file per process. A mini-application is defined to insert random 64-bit \texttt{integer} numbers into the DHT. Each value is inserted by calculating the LV owner through a hash function that is designed to guarantee that all the LVs are equally filled. The number of inserts per test depends on the overall size of the DHT, which is determined by {\ttfamily num\_procs} $\times$ ({\ttfamily local\_volume} + {\ttfamily heap\_factor} $\times$ {\ttfamily local\_volume}). The overflow heap factor is set to 4, so that four extra elements will be allocated in the heap per element inside the LV.

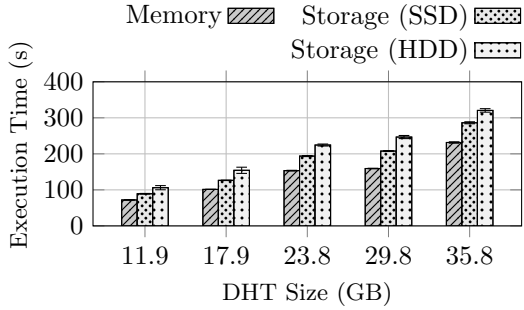
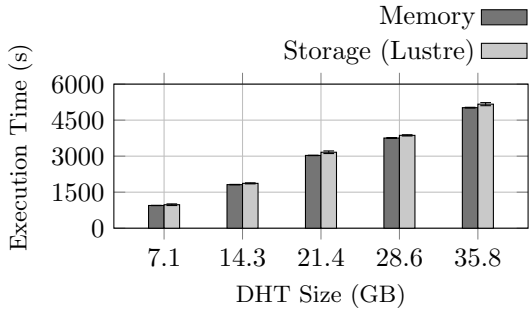
\begin{figure}
    \centering
    \begin{subfigure}[t]{0.4921\textwidth}
        \centering
        \begin{tikzpicture}
            \begin{axis}[
                xlabel=DHT Size (GB),
                ylabel=Execution Time (s),
                symbolic x coords={11.9,17.9,23.8,29.8,35.8},
                xtick=data,
                ytick={0,100,200,300,400},
                ymin=0,
                ymax=400,
                scaled y ticks = false,
                y tick label style={/pgf/number format/fixed, /pgf/number format/1000 sep = },
                ylabel style={at={(0.05,0.5621)}, style={font=\small}},
                xlabel style={at={(0.5,-0.05)}, style={font=\small}},
                enlarge x limits=0.16,
                grid=major,
                legend style={legend columns=2,at={(1.0,1.581)},anchor=north east},
                legend style={/tikz/every even column/.append style={column sep=4.21}},
                legend cell align=right,
                legend plot pos=right,
                legend style={draw=none, fill=white, inner xsep=0, inner ysep=0},
                ybar=0pt,
                bar width=5.81pt,
                width=0.8921\textwidth,
                height=3.4921cm,
                area legend
            ]
            \addplot [fill=black!21, postaction={pattern=north east lines}] plot[error bars/.cd, y dir=both, y explicit] table [y error minus=min, y error plus=max] {
                x       y       min     max
                11.9   72.03	0.62    0.62
                17.9   101.34	0.12    0.12
                23.8   153.45	0.89    0.89
                29.8   159.30	0.68    0.68
                35.8   231.53	1.95    1.95
            };\addlegendentry{Memory}
            \addplot [fill=black!9, postaction={pattern=crosshatch dots}] plot[error bars/.cd, y dir=both, y explicit] table [y error minus=min, y error plus=max] {
                x       y       min     max
                11.9   89.21   0.91    0.91
                17.9   126.44  0.59    0.59
                23.8   193.80  1.40    1.40
                29.8   207.79  1.29    1.29
                35.8   286.27  2.74    2.74
            };\addlegendentry{Storage (SSD)}
            
            \addlegendimage{empty legend}
            \addlegendentry{}
            
            \addplot [fill=black!5, postaction={pattern=dots}] plot[error bars/.cd, y dir=both, y explicit] table [y error minus=min, y error plus=max] {
                x       y       min     max
                11.9   106.05  5.80    5.80
                17.9   154.30  8.61    8.61
                23.8   224.48  2.77    2.77
                29.8   246.64  4.32    4.32
                35.8   320.57  4.89    4.89
            };\addlegendentry{Storage (HDD)}
            \end{axis}
        \end{tikzpicture}
        \vspace{-0.21cm}
        \caption{\textbf{Blackdog Results} / 8 MPI Processes} \label{fig:dht_blackdog}
    \end{subfigure}
    
    \vspace{0.3921cm}
    
    \begin{subfigure}[t]{0.47921\textwidth}
        \centering
        \begin{tikzpicture}
            \begin{axis}[
                xlabel=DHT Size (GB),
                ylabel=Execution Time (s),
                symbolic x coords={7.1,14.3,21.4,28.6,35.8},
                xtick=data,
                ytick={0,1500,3000,4500,6000},
                ymin=0,
                ymax=6000,
                scaled y ticks = false,
                y tick label style={/pgf/number format/fixed, /pgf/number format/1000 sep = },
                ylabel style={at={(0.0,0.5621)}, style={font=\small}},
                xlabel style={at={(0.5,-0.05)}, style={font=\small}},
                enlarge x limits=0.16,
                grid=major,
                legend style={legend columns=1,at={(1.0,1.581)},anchor=north east},
                legend style={/tikz/every even column/.append style={column sep=4.21}},
                legend cell align=right,
                legend plot pos=right,
                legend style={draw=none, fill=white, inner xsep=0, inner ysep=0},
                ybar=0pt,
                bar width=5.81pt,
                width=0.8921\textwidth,
                height=3.4921cm,
                area legend
            ]
            \addplot [fill=black!54] plot[error bars/.cd, y dir=both, y explicit] table [y error minus=min, y error plus=max] {
                x       y           min     max
                7.1    946.12      5.84    5.84
                14.3   1814.99     9.85    9.85
                21.4   3030.74     8.50    8.50
                28.6   3755.73     19.45   19.45
                35.8   5019.42     16.38   16.38
            };\addlegendentry{Memory}
            \addplot [fill=black!21] plot[error bars/.cd, y dir=both, y explicit] table [y error minus=min, y error plus=max] {
                x       y           min     max
                7.1    978.12      34.58   34.58
                14.3   1867.80     21.82   21.82
                21.4   3163.38     54.64   54.64
                28.6   3866.35     26.39   26.39
                35.8   5166.20     64.45   64.45
            };\addlegendentry{Storage (Lustre)}
            \end{axis}
        \end{tikzpicture}
        \vspace{-0.21cm}
        \caption{\textbf{Tegner Results} / 96 MPI Processes} \label{fig:dht_tegner}
    \end{subfigure}
    \caption{Distributed Hash Table performance using MPI memory windows and MPI storage windows on Blackdog and Tegner. Note the differences in process count between each testbed.\label{fig:dht_blackdog_tegner}}
\end{figure}

The use of atomic operations inside the DHT implementation effectively hides some of the constraints expected due to the memory and storage bandwidth differences. \autoref{fig:dht_blackdog} presents the average execution time on Blackdog using MPI memory windows and MPI storage windows. We use 8 MPI processes and each process inserts 80\% of the LV capacity. This means that, in total, 80\% of the DHT will be filled. The tests vary the DHT size from 11.92GB (20 million elements per LV) up to 35.76GB (60 million elements per LV). The results show that the overhead of using MPI storage windows with conventional hard disks is approximately 32\% on average compared to using MPI memory windows. Using the SSD, the performance improves by decreasing the overhead to approximately 20\% on average. Even though the mini-application is designed to perform mostly write operations, the page cache is effectively hiding the bandwidth differences by aggregating most of these write operations. We estimate that better storage technologies (e.g., NVRAM) should approximate the performance to that of the memory-based implementation.

By increasing the process count and the number of active nodes, the overhead of the network communication and atomic CAS operations required to maintain the DHT increases due to LV insert conflicts. Hence, hiding the performance limitations. \autoref{fig:dht_tegner} shows the average execution time on Tegner using MPI memory windows and MPI storage windows mapped into a Lustre parallel file system. These tests use 96 processes on 4 compute nodes. Each process inserts 80\% of the LV capacity to the table. The LV size varies from 7.15GB (1 million elements per LV) to 35.76GB (5 million elements per LV). Note that the number of elements per LV differs from the previous test, as the process count has considerably increased. In this case, we observe that using MPI storage windows barely affects the performance with only a 2\% degradation on average when compared to MPI memory windows. The execution times are clearly dominated by the use of atomic operations to resolve the LV / heap conflicts\footnote{Atomic operations on Tegner reduce the sustained bandwidth in half (\autoref{fig:imb_getaccumulate}).}.

\subsection{Out-of-Core Computation}

A large number of HPC applications have to deal with very large datasets that exceed the main memory capacity. In such cases, out-of-core techniques~\cite{thakur1996extended, cox1997application} define efficient mechanisms to transfer data from / to storage. For instance, a common approach is to divide the main algorithm in blocks~\cite{toledo1999survey}. This is particularly useful in certain applications that involve dense matrix computations. However, the programmer is responsible for the distribution and the associated data transferring, which introduces source code complexity.

MPI storage windows provide a seamless extension to the main memory by mapping part of the storage into the memory space. This enables applications to transparently use more memory than physically available, without the burden of handling any data management. In addition, applications can opt to use combined window allocations through the ``factor'' performance hint. This hint provides a mechanism that allows applications to define a combination of traditional MPI memory windows and MPI storage windows, while still maintaining consecutive memory addresses. Hence, applications can ensure that only the part of the memory that exceeds the main memory limit is mapped to storage, avoiding the OS to interfere with the rest of the allocation\footnote{The statement is only valid if no swap mechanism exists, as in our testbeds. Hence, allowing us to avoid interferences on the allocated memory region. Future implementations could also move large portions of memory from / to storage to increase the performance.}.

In this regard, we evaluate the performance of MPI storage windows and combined window allocations using the DHT implementation that was presented in the previous subsection. We aim to understand the implications of out-of-core on real-world HPC applications. The LV and the overflow heap are allocated as MPI storage windows on each process. All the updates to the DHT are handled using solely MPI one-sided operations. In addition, we instruct the mini-application to increase the overall capacity of the DHT up to almost twice times the main memory capacity. This should give us an idea of what would be the overhead of exceeding the physical memory limit.

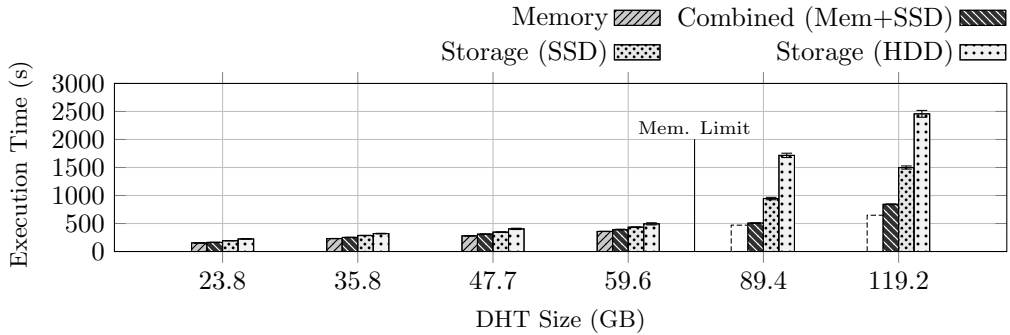
\begin{figure*}
    \centering
    \hspace{-0.81cm}
    \begin{tikzpicture}
        \begin{axis}[
            xlabel=DHT Size (GB),
            ylabel=Execution Time (s),
            symbolic x coords={23.8,35.8,47.7,59.6,89.4,119.2},
            xtick=data,
            ytick={0,500,1000,1500,2000,2500,3000},
            ymin=0,
            ymax=3000,
            scaled y ticks = false,
            y tick label style={/pgf/number format/fixed, /pgf/number format/1000 sep = },
            ylabel style={at={(0.0,0.521)}, style={font=\small}},
            xlabel style={at={(0.5,-0.05)}, style={font=\small}},
            enlarge x limits=0.16,
            grid=major,
            legend style={legend columns=2,at={(1.0,1.496)},anchor=north east},
            legend style={/tikz/every even column/.append style={column sep=4.21}},
            legend cell align=right,
            legend plot pos=right,
            legend style={draw=none, fill=white, inner xsep=0, inner ysep=0},
            ybar=0pt,
            bar width=5.81pt,
            width=0.81\textwidth,
            height=3.81cm,
            area legend
        ]
        
        \draw [dash pattern={on 2pt off 1pt}, draw=black, fill=white]  (6.6921cm,0cm) rectangle (6.921cm,0.34921cm);
        \draw [dash pattern={on 2pt off 1pt}, draw=black, fill=white]  (8.481cm,0cm) rectangle (9.21cm,0.481cm);

        \draw  (6.21cm,0) |- (6.21cm,1.49cm) [solid] node [above, fill=white, inner xsep=1.81, inner ysep=1.921] {\scriptsize{Mem. Limit}};
        
        \addplot [fill=black!21, postaction={pattern=north east lines}] plot[error bars/.cd, y dir=both, y explicit] table [y error minus=min, y error plus=max] {
            x       y       min     max
            23.8    153.45  0.89    0.89
            35.8    231.53  1.95    1.95
            47.7    279.55  1.34    1.34
            59.6    358.33  1.91    1.91
            89.4    0       0       0
            119.2   0       0       0
        };\addlegendentry{Memory}
        \addplot [fill=black!81, postaction={pattern=north west lines, pattern color=white}] plot[error bars/.cd, y dir=both, y explicit] table [y error minus=min, y error plus=max] {
            x       y           min             max
            23.8    162.923745  1.5927096212    1.5927096212
            35.8    251.3134172 3.8679585327    3.8679585327
            47.7    309.9640156 4.4099900998    4.4099900998
            59.6    391.8037598 6.4184560946    6.4184560946
            89.4    506.6676488 3.7293788526    3.7293788526
            119.2   842.3803338 6.3409614888    6.3409614888
        };\addlegendentry{Combined (Mem+SSD)}
        \addplot [fill=black!9, postaction={pattern=crosshatch dots}] plot[error bars/.cd, y dir=both, y explicit] table [y error minus=min, y error plus=max] {
            x       y       min     max
            23.8    193.80  1.40    1.40
            35.8    286.27  2.74    2.74
            47.7    346.30  5.06    5.06
            59.6    434.31  5.47    5.47
            89.4    945.75  21.62   21.62
            119.2   1499.29 27.78   27.78
        };\addlegendentry{Storage (SSD)}
        \addplot [fill=black!5, postaction={pattern=dots}] plot[error bars/.cd, y dir=both, y explicit] table [y error minus=min, y error plus=max] {
            x       y       min     max
            23.8    224.48  2.77    2.77
            35.8    320.57  4.89    4.89
            47.7    404.10  11.39   11.39
            59.6    493.89  18.76   18.76
            89.4    1716.02 38.81   38.81
            119.2   2457.47 61.05   61.05
        };\addlegendentry{Storage (HDD)}
        \end{axis}
    \end{tikzpicture}
    \caption{Distributed Hash Table performance of MPI memory windows, MPI storage windows and combined window allocations on Blackdog (8 MPI Processes). The dashed bars indicate the projected value using MPI memory windows after the main memory is exceeded.\label{fig:ooc_dht_blackdog}}
\end{figure*}

With a small performance overhead, we observe that using MPI storage on applications that exceed the main memory capacity is feasible. \autoref{fig:ooc_dht_blackdog} shows the average execution time on Blackdog using MPI memory windows, MPI storage windows, and combined window allocations. We use 8 MPI processes and each process inserts 80\% of the LV capacity to the table, as in our previous tests. We vary the DHT size from 23.84GB (40 million elements per LV) up to 119.21GB (200 million elements per LV). With MPI memory windows, we observe that the mini-application cannot continue executing in the last two cases. This is due to the physical memory limit of Blackdog, which amounts to 72GB of DRAM. Nonetheless, by using MPI storage windows, the execution can proceed without any changes in the source code. The performance penalty with conventional hard disks before exceeding the main memory limit is approximately 32\% compared to MPI memory windows. The overhead of using SSD is approximately 20\% on average. After exceeding the main memory limit, however, the performance penalty considerably increases. Using the local hard disk, the overhead is over twice as much as the projected value using MPI memory windows. With the SSD, the overhead increases to approximately 89\%. This is mainly due to the fact that our tests enforce mostly write operations with no data reuse. Thus, the performance depends on the storage bandwidth.

On the other hand, we observe that combined window allocations can provide the performance benefits of using MPI memory windows with the versatility of MPI storage windows for out-of-core execution. \autoref{fig:ooc_dht_blackdog} illustrates the performance of using combined window allocations with a fixed factor of \texttt{0.5}. This means that 50\% of the allocation is located in memory, while the other 50\% is located on storage. The SSD is used for the storage mapping in our tests. Before exceeding the main memory limit, we observe an average overhead of only 8\% when compared to MPI memory windows. After exceeding the memory capacity, the overhead is increased to 13\% in comparison with the projected value of MPI memory windows. In the largest test case, where 59.6GB of the allocation is based on storage, the overhead increases to only 36\% on average. The main motivation behind this excellent result is due to the fact that not every consecutive byte of the allocation is mapped to storage. The part of the allocation that resides on storage is mainly the heap designated for conflicts. Hence, the application predominantly pays only the cost of insert conflicts during our tests. The rest of the operations hit mainly in the LV of each process, which is mostly located (pinned) in memory.

\subsection{Parallel I/O}

In this subsection, we briefly evaluate the performance of MPI storage windows in comparison with individual and collective I/O operations of MPI I/O.

\subsubsection{Checkpoint-Restart on HACC I/O Kernel}

The MPI standard offers support for high-performance parallel I/O through the MPI I/O specification~\cite{MPI3standard2015}. Optimizations such as data sieving or two-phase I/O~\cite{thakur1999data, thakur1997users}, included in most of the popular MPI implementations, provide collective access to the underlying file system. Moreover, MPI I/O allows for non-contiguous access to storage~\cite{ching2003noncontiguous}, effectively increasing the overall throughput of HPC applications compared to other alternatives (e.g., POSIX I/O).

\begin{figure}
    \begin{subfigure}[t]{0.47921\textwidth}
        \centering
        \begin{tikzpicture}
            \begin{axis}[
                xlabel=MPI Processes,
                ylabel=Execution Time (s),
                symbolic x coords={2,4,8,16},
                xtick=data,
                ytick={0,10,20,30,40},
                ymin=0,
                ymax=40,
                scaled y ticks = false,
                y tick label style={/pgf/number format/fixed, /pgf/number format/1000 sep = },
                ylabel style={at={(0.0921,0.4921)}, style={font=\small}},
                xlabel style={at={(0.5,-0.05)}, style={font=\small}},
                enlarge x limits=0.21,
                grid=major,
                legend style={legend columns=1,at={(1.0,1.581)},anchor=north east},
                legend style={/tikz/every even column/.append style={column sep=4.21}},
                legend cell align=right,
                legend plot pos=right,
                legend style={draw=none, fill=white, inner xsep=0, inner ysep=0},
                ybar=0pt,
                bar width=5.81pt,
                width=0.8921\textwidth,
                height=3.4921cm,
                area legend
            ]
            \addplot [fill=black!21, postaction={pattern=dots}] plot[error bars/.cd, y dir=both, y explicit] table [y error minus=min, y error plus=max] {
                x       y       min     max
                2       28.91   0.57    0.57
                4       27.23   0.14    0.14
                8       26.61   0.16    0.16
                16      28.45   0.15    0.15
            };\addlegendentry{MPI I/O (HDD)}
            \addplot [fill=black!5, postaction={pattern=dots}] plot[error bars/.cd, y dir=both, y explicit] table [y error minus=min, y error plus=max] {
                x       y       min     max
                2       30.08   0.16    0.16
                4       28.55   0.12    0.12
                8       28.35   0.11    0.11
                16      29.19   0.07    0.07
            };\addlegendentry{Storage Win. (HDD)}
            \end{axis}
        \end{tikzpicture}
        \vspace{-0.21cm}
        \caption{\textbf{Blackdog Results} / 100M Particles} \label{fig:hacc_blackdog}
    \end{subfigure}
    
    \vspace{0.3921cm}
    
    \begin{subfigure}[t]{0.47921\textwidth}
        \centering
        \hspace{0.021cm}
        \begin{tikzpicture}
            \begin{axis}[
                xlabel=MPI Processes,
                ylabel=Execution Time (s),
                symbolic x coords={16,32,64,128},
                xtick=data,
                ytick={0,3,6,9,12},
                ymin=0,
                ymax=12,
                scaled y ticks = false,
                y tick label style={/pgf/number format/fixed, /pgf/number format/1000 sep = },
                ylabel style={at={(0.0921,0.4921)}, style={font=\small}},
                xlabel style={at={(0.5,-0.05)}, style={font=\small}},
                enlarge x limits=0.21,
                grid=major,
                legend style={legend columns=1,at={(1.0,1.581)},anchor=north east},
                legend style={/tikz/every even column/.append style={column sep=4.21}},
                legend cell align=right,
                legend plot pos=right,
                legend style={draw=none, fill=white, inner xsep=0, inner ysep=0},
                ybar=0pt,
                bar width=5.81pt,
                width=0.8921\textwidth,
                height=3.4921cm,
                area legend
            ]
            \addplot [fill=black!39] plot[error bars/.cd, y dir=both, y explicit] table [y error minus=min, y error plus=max] {
                x       y       min     max
                16      5.90    0.28    0.28
                32      7.25    0.64    0.64
                64      8.63    0.96    0.96
                128     9.89    0.81    0.81
            };\addlegendentry{MPI I/O (Lustre)}
            \addplot [fill=black!21] plot[error bars/.cd, y dir=both, y explicit] table [y error minus=min, y error plus=max] {
                x       y       min     max
                16      8.05    0.58    0.58
                32      5.38    0.23    0.23
                64      6.59    0.21    0.21
                128     7.63    0.29    0.29
            };\addlegendentry{Storage Win. (Lustre)}
            \end{axis}
        \end{tikzpicture}
        \vspace{-0.21cm}
        \caption{\textbf{Tegner Results} / 100M Particles} \label{fig:hacc_tegner}
    \end{subfigure}
    \caption{HACC I/O performance using MPI I/O and MPI storage windows running on Blackdog and Tegner.\label{fig:hacc_blackdog_tegner}}
\end{figure}
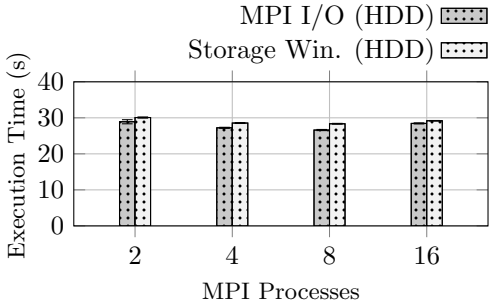
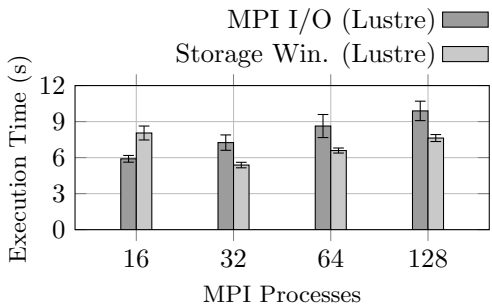

MPI storage windows can also be used as a very simple API to perform individual I/O operations. Without setting aside the advantages of MPI I/O, the approach provides a unified programming interface that streamlines the accesses to the storage layer on already existing one-sided applications. In this way, programmers can use \texttt{put} and \texttt{get} operations as an alternative to the \texttt{write} and \texttt{read} operations of MPI I/O. In addition, \texttt{load} / \texttt{store} operations over the local pointer of the window are possible. Thus, MPI storage windows simplifies I/O handling while providing advantages, such as caching.

We use the checkpoint-restart mechanism of HACC (Hardware / Hybrid Accelerated Cosmology Code)~\cite{habib2016hacc} to evaluate the I/O capabilities of MPI storage windows. HACC is a physics particle-based code to simulate the evolution of the Universe after the Big Bang that operates on the trajectories of trillions of particles. For our tests, we evaluate the performance using the HACC I/O kernel, a mini-application of the CORAL benchmark codes\footnote{\url{https://asc.llnl.gov/CORAL-benchmarks}} that mimics the checkpoint-restart functionality of the original code. We extend the kernel to use MPI storage windows during this process and compare it with the performance of the existing MPI I/O implementation. This latter implementation uses individual file I/O (i.e., not collective operations). The particle data is stored inside a global shared file. For fair comparison, we also ensure a synchronization point during checkpoint to avoid any buffering / caching. 

We determine that, with slight performance differences, MPI storage windows can be used as an alternative to MPI I/O when using local storage and individual I/O operations. \autoref{fig:hacc_blackdog} shows the average execution time using MPI I/O and MPI storage windows on Blackdog, both targeting a local hard disk. We use 100 million particles in all the evaluations while doubling the number of processes (i.e., strong scaling). From the results, we observe that the performance of the two approaches is similar on Blackdog. MPI I/O performs slightly better on average, being between a 2-4\% faster in most cases. We assume that the MPI storage windows version introduces a slight overhead due to the inherent page faults, that trigger the accesses to storage. The code does not seem to scale on either version.

On the other hand, increasing the process count can be beneficial for MPI storage windows in some situations. \autoref{fig:hacc_tegner} shows the same strong scaling evaluation on Tegner, using remote storage through Lustre. In this case, MPI storage windows provide about a 32\% improvement on average when compared to the MPI I/O implementation. We observe this improvement mostly after the process count increases from 16 to 32 processes, which also raises the number of nodes in use from one to two active nodes. This result indicates that, in some cases, MPI storage windows might provide better scalability compared to using individual operations of MPI I/O. Nonetheless, we assume that collective I/O operations would be more beneficial in this case to access the shared file. These operations aggregate multiple I/O requests through dedicated I/O processes. Hence, contention can be reduced on parallel file systems (e.g., Lustre).

\subsubsection{Transparent Checkpointing}

Over the past few years, resilience has become one of the major concerns in HPC~\cite{cappello2009toward}. With the arrival of the first wave of pre-Exascale machines, the chances for unexpected failures during the execution of parallel applications will considerably increase~\cite{bland2012user}. Hence, several solutions have been proposed to mitigate failures at user-level~\cite{gropp2004fault, bland2012proposal, guo2015fault, fagg2000ft}. The importance of these solutions might even affect the design decisions of the upcoming revisions of the MPI standard. As a consequence, we observe the need for efficient resilience support on current and upcoming HPC clusters.

In this regard, MPI storage windows can be used to provide user-level fault-tolerance support. By transparently integrating storage into the memory management of HPC applications, the approach offers a very simple yet efficient method to define novel mechanisms that protect against failures. For instance, data transferring from / to storage is overlapped with computations. This means that only certain synchronization points with the storage layer (i.e., using \texttt{MPI\_Win\_sync}) are required to maintain data consistency.

\input{3_Results_6_FaultTolerance_fig1}

For this purpose, we present a checkpoint mechanism inside \mbox{MapReduce-1S}\footnote{\url{https://github.com/sergiorg-kth/mpi-mapreduce-1s}} (MapReduce ``One-Sided''). This project is an on-going effort that proposes the integration of a decentralized strategy for MapReduce frameworks using MPI one-sided communication to overlap the execution of the Map and Reduce phases of the algorithm. The aim is to decrease the workload imbalance across the processes on input datasets with irregular distribution of the data. 
The implementation uses a complex multi-window configuration and non-blocking I/O to reduce the overhead while reading the input datasets.

We introduce support for MPI storage windows and extend MapReduce-1S to perform a window synchronization point after each Map task, as well as after the Reduce phase is completed. In addition, we evaluate the checkpoint performance in comparison with a reference implementation by Hoefler et al.~\cite{hoefler2009towards}, that employs state-of-the-art collective communication and I/O. We extend this implementation to perform each checkpoint using MPI collective I/O over a shared file. Our goal is to evaluate how MPI storage windows could reduce the latency of checkpointing compared to a traditional solution with MPI I/O. The evaluations are conducted using Word-Count on a large dataset from the Purdue MapReduce Benchmarks Suite (PUMA)~\cite{ahmad2012puma}. In particular, we use the \emph{Dataset3} from the PUMA-Wikipedia datasets\footnote{\url{https://engineering.purdue.edu/~puma/datasets.htm}}. We pre-process the files off-line to generate unified, large input datasets that guarantee equivalent workload per process. Note that the labels MR-2S (MapReduce ``Two-Sided'') and MR-1S are used to refer to each implementation.

Being able to combine computations with storage operations is clearly one of the main advantages of MPI storage windows. \autoref{fig:benchmarks_checkpoint_strong} illustrates the performance of MR-2S and MR-1S by varying the number of MPI processes on Tegner for a fixed-size input dataset (strong scaling), with checkpoint support. For reference purposes, the label ``[NoFT]'' (i.e., No Fault-Tolerance) illustrates the baseline implementation performance without checkpoint support. The process count varies from 64 (3 nodes) up to 512 (22 nodes). We use an input dataset from PUMA-Wikipedia with 32GB of data, and a task size of 64MB per process. Thus, the number of checkpoints varies from 8 (64 processes) down to 1 (512 processes). Despite that the overall performance of MR-2S without checkpoint support is 9.6\% faster than MR-1S on average, we determine from this figure that adding checkpoint support affects the scalability on higher process counts. For instance, using 512 processes, MR-1S with checkpoint support is up to 17.6\% faster in comparison with MR-2S. In this particular case, the checkpoint overhead is 21.2\% using MPI storage windows on MR-1S, and 58.6\% using collective MPI I/O operations on MR-2S. This observation is depicted on \autoref{fig:benchmarks_checkpoint_strong_overhead}.

By increasing the size of the input datasets and, consequently, the workload per process, we confirm that MPI storage windows provide advantages for fault-tolerance. \autoref{fig:benchmarks_checkpoint_weak} shows the performance of MR-2S and MR-1S by varying the number of MPI processes on Tegner and maintaining the workload per process (weak scaling), with and without checkpoint support. The process count varies from 64 (3 nodes) up to 512 (22 nodes). We use reference input datasets from PUMA-Wikipedia, with a fixed 1GB workload per process (i.e., input sizes from 64GB to 512GB). The number of checkpoints is also fixed to 16 per test. From this figure, we determine that MR-2S is 6.0\% faster than MR-1S on average when no checkpoints are conducted. However, as the process count increases, we confirm that the performance of MR-1S with checkpoint support is 59.3\% faster in the last case of 512 processes, in comparison with MR-2S. \autoref{fig:benchmarks_checkpoint_weak_overhead} reflects the checkpoint overhead per implementation. Using MPI storage windows on MR-1S only incurs in a 3.8\% penalty on average.

\section{Discussion}
\label{4_Discussion}

The performance results given in the previous section have illustrated some of the benefits of using MPI storage windows. Here we extend the discussion concerning these results.

\subsubsection*{Limitations of memory-mapped IO}

Introducing storage operations as part of the memory space of an application allows us to combine computations and storage operations, hiding part of the bandwidth and access latency differences between memory and storage. However, we note that the use of memory-mapped I/O for MPI storage windows might constraint the performance on large-process counts~\cite{pumma2017parallel}. This is due to the inherent context-switches required when multiple processes are simultaneously conducting I/O and must wait for the I/O requests to succeed. In addition, we also observe that applications can inevitably be bounded by the storage bandwidth when performing only large, irregular \texttt{write} operations over the mapped storage space. Even if the page cache of the OS could theoretically absorb part of these modifications, the changes must be synchronized with storage at some point and might produce stall periods on the process. We expect that future implementations based on MPI I/O can avoid some of these limitations by transferring large portions of memory from / to storage.

\subsubsection*{Combining memory and storage allocations}

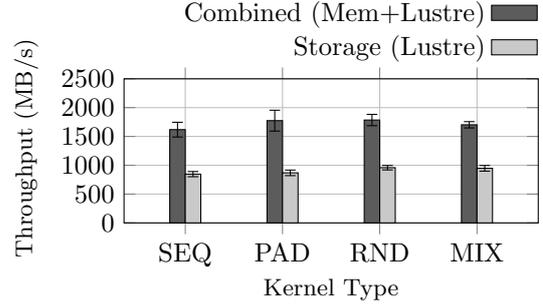
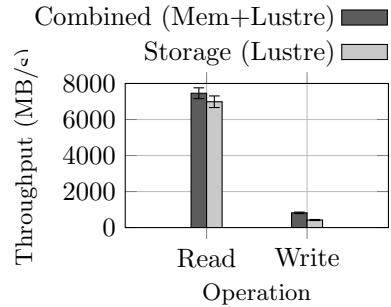
\begin{figure}
    \begin{subfigure}[t]{0.47921\textwidth}
        \centering
        \begin{tikzpicture}
            \begin{axis}[
                xlabel=Kernel Type,
                ylabel=Throughput (MB/s),
                symbolic x coords={SEQ, PAD, RND, MIX},
                xtick=data,
                ytick={0,500,1000,1500,2000,2500},
                ymin=0,
                ymax=2500,
                scaled y ticks = false,
                y tick label style={/pgf/number format/fixed, /pgf/number format/1000 sep = },
                ylabel style={at={(-0.01,0.521)}, style={font=\small}},
                xlabel style={at={(0.5,-0.05)}, style={font=\small}},
                enlarge x limits=0.21,
                grid=major,
                legend style={legend columns=1,at={(1.0,1.581)},anchor=north east},
                legend style={/tikz/every even column/.append style={column sep=4.21}},
                legend cell align=right,
                legend plot pos=right,
                legend style={draw=none, fill=white, inner xsep=0, inner ysep=0},
                ybar=0pt,
                bar width=5.81pt,
                width=0.8921\textwidth,
                height=3.4921cm,
                area legend
            ]
            \addplot [fill=black!64] plot[error bars/.cd, y dir=both, y explicit] table [y error minus=min, y error plus=max] {
                x       y               min             max
                SEQ     1618.099799     127.6748607399  127.6748607399
                PAD     1774.0558832    181.0064829277  181.0064829277
                RND     1784.7435708    98.0106432673   98.0106432673
                MIX     1702.2941244    54.3283056579   54.3283056579
            };\addlegendentry{Combined (Mem+Lustre)}
            \addplot [fill=black!21] plot[error bars/.cd, y dir=both, y explicit] table [y error minus=min, y error plus=max] {
                x       y               min             max
                SEQ     846.2173242     45.0242460873   45.0242460873
                PAD     868.5976124     48.4561556867   48.4561556867
                RND     958.138817      38.7252254353   38.7252254353
                MIX     947.0190864     48.8732941296   48.8732941296
            };\addlegendentry{Storage (Lustre)}
            \end{axis}
        \end{tikzpicture}
        \vspace{-0.21cm}
        \caption{\textbf{Tegner Results} / 1 MPI Process} \label{fig:discussion_benchmark_tegner}
    \end{subfigure}
    
    \vspace{0.3921cm}
    
    \begin{subfigure}[t]{0.47921\textwidth}
        \centering
        \begin{tikzpicture}
            \begin{axis}[
                xlabel=Operation,
                ylabel=Throughput (MB/s),
                symbolic x coords={Read, Write},
                xtick=data,
                ytick={0,2000,4000,6000,8000},
                ymin=0,
                ymax=8000,
                scaled y ticks = false,
                y tick label style={/pgf/number format/fixed, /pgf/number format/1000 sep = },
                ylabel style={at={(-0.021,0.4921)}, style={font=\small}},
                xlabel style={at={(0.5,-0.05)}, style={font=\small}},
                enlarge x limits=0.81,
                grid=major,
                legend style={legend columns=1,at={(1.0,1.581)},anchor=north east},
                legend style={/tikz/every even column/.append style={column sep=4.21}},
                legend cell align=right,
                legend plot pos=right,
                legend style={draw=none, fill=white, inner xsep=0, inner ysep=0},
                ybar=0pt,
                bar width=5.81pt,
                width=0.64\textwidth,
                height=3.4921cm,
                area legend
            ]
            \addplot [fill=black!64] plot[error bars/.cd, y dir=both, y explicit] table [y error minus=min, y error plus=max] {
                x       y               min             max
                Read    7457.8555526    300.9855701946  300.9855701946
                Write   817.2108746     42.2792216599   42.2792216599
            };\addlegendentry{Combined (Mem+Lustre)}
            \addplot [fill=black!21] plot[error bars/.cd, y dir=both, y explicit] table [y error minus=min, y error plus=max] {
                x       y               min             max
                Read    6986.2281294    322.2979302789  322.2979302789
                Write   419.2369286     29.9754495131   29.9754495131
            };\addlegendentry{Storage (Lustre)}
            \end{axis}
        \end{tikzpicture}
        \vspace{-0.21cm}
        \caption{\textbf{Avg. Throughput} on Tegner} \label{fig:discussion_benchmark_rw_tegner}
    \end{subfigure}
    \caption{(\subref*{fig:discussion_benchmark_tegner}) mSTREAM microbenchmark performance using MPI storage windows and combined window allocations on Tegner. (\subref*{fig:discussion_benchmark_rw_tegner}) Read and write performance evaluation using the \texttt{SEQ} kernel with MPI storage windows and combined window allocations on Tegner.\label{fig:discussion_tegner}}
\end{figure}

Combined window allocations divide a range of consecutive virtual addresses into memory and storage. Two separate mappings are then established with a fixed subset of the reserved virtual addresses. One of the main benefits of this type of allocations is that they reduce the overhead of letting the OS manage the full allocation. Thus, the memory part is inherently ``pinned''. \autoref{fig:discussion_tegner} demonstrates that the approach can also benefit clusters that use Lustre, such as Tegner. The example uses a fixed factor of \texttt{0.5} to simulate that half of the allocation cannot fit into memory. Despite the observed differences in \texttt{read} / \texttt{write} performance, the figure illustrates that the throughput increases up to 1.68GB/s on average, almost twice as much compared to the original results.

\subsubsection*{Going beyond the main memory limit}

In some situations, finding a good balance between computations and \texttt{read} / \texttt{write} operations is not possible. A good example is going beyond the physical memory limit while performing write operations (see \autoref{fig:ooc_dht_blackdog}). After exceeding the main memory limit, the throughput strictly depends on the transfer rate of the storage device used. Nonetheless, this trade-off compensates the fact that traditional MPI memory windows cannot exceed the \texttt{vm.overcommit\_ratio}. This limit determines the maximum amount of memory that a certain application is allowed to allocate (in our tests, 90\% of the main memory). Without increasing the physical memory capacity or using techniques such as out-of-core, the execution of any HPC application will fail after reaching this limit. Hence, using MPI storage windows can be beneficial by transparently hiding the complexity of managing these situations.

\subsubsection*{Buffering on Lustre for better performance}
We observed that MPI storage windows did not perform well in some experiments that use the Lustre parallel file system mounted on Tegner. We demonstrated that this effect was due to the asymmetric \texttt{read} / \texttt{write} performance featured on the cluster. The reason for the asymmetry is related to how the Lustre client of each node handles internally the memory-mapped I/O. For instance, regular I/O may benefit from increasing the amount of dirty client cache through the \texttt{max\_dirty\_mb} setting of Lustre. Unfortunately, memory-mapped file I/O or regular I/O that uses the \texttt{O\_DIRECT} flag, are configured to cache data up to a full RPC. At this point, all the written data is effectively transferred to the correspondent OST server. We expect to design and implement a custom implementation that resembles the memory-mapped I/O mechanism at user-level. This will offer full-control of the mapping from the MPI implementation.

\subsubsection*{Transparent checkpoint support}

MPI storage windows can provide advantages for novel fault-tolerance mechanisms. By just relying on synchronization points that include an \texttt{MPI\_Win\_sync} call over the window object, we illustrated how checkpoint support can be implemented to ensure data consistency with the storage layer. In this regard, we note that decoupled checkpointing can be possible by using exclusive locks over the local MPI storage window. \autoref{exampleCodeFaultTolerance} highlights how easily one could ensure data consistency while preventing remote processes to access the local information that is exposed through the window. This avoids the use of a global \texttt{MPI\_Barrier}. Lastly, we observe that version control after each checkpoint can be easily maintained at application-level. We can understand the status of the flushed data by including a header alongside individual or grouped values~\cite{huansongkeyval2017}. Alternatively, a simple approach is to use two MPI storage windows and swap them on each checkpoint. Hence, we guarantee that the OS will only flush the modified data on the active window.

\begin{lstlisting}[caption={Synchronization point that guarantees data consistency with storage.}, label={exampleCodeFaultTolerance}]
...
void checkpoint()
{
    // Lock window to prevent changes
    MPI_Win_lock(MPI_LOCK_EXCLUSIVE,
                 myrank, 0, win_keyval);
    MPI_Win_sync(win_keyval);
    MPI_Win_unlock(myrank, win_keyval);
}
...
\end{lstlisting}

\subsubsection*{Performance considerations with MPI I/O}



In certain use-cases, such as during checkpoint in MapReduce-1S, we observed that MPI storage windows can provide advantages over MPI I/O. For instance, the selective synchronization mechanism of memory-mapped IO decreases the checkpoint overhead by avoiding to flush all the data from memory to storage. Hence, a call to \texttt{MPI\_Win\_sync} only ensures that the recently updated data inside the window is correctly stored in the correspondent mapped file. By using MPI I/O, however, each checkpoint requires to flush the current status of each process into the global shared file. Thus, unless applications integrate their own mechanism to keep track of the modified data, the performance is compromised in comparison. This is the main reason why the use of collective I/O did not provide any additional advantages, despite this type of operations generally providing higher performance in several orders of magnitude~\cite{gropp2014using}. Nonetheless, we also note that a combination of non-blocking MPI I/O operations and local buffering might provide similar benefits, even when page caching is supported.


\section{Related Work}
\label{5_RelatedWork}
MPI windows have been included in MPI since MPI-2 through the one-sided communication model with the main purpose of utilizing RDMA for lower communication overhead compared to explicit message passing~\cite{gropp2014using}. MPI-3 further extends the RMA interface and~\cite{gerstenberger2014enabling} provides a high-performance implementation. The previous work~\cite{dorozynski2016checkpointing} has proposed using MPI windows to support checkpointing on emerging non-volatile memories. In contrast, our work extends the concept of MPI window to a broad spectrum of storage devices. MPI storage windows allow for interaction with different tiers of the storage hierarchy, providing a portable I/O mechanism. Our approach also enables seamless data transferring from memory to storage by utilizing the highly-optimized OS support for paging.

Large-scale scientific applications often suffer from I/O operations as their performance bottleneck. In addition, as data-intensive applications are emerging on supercomputers, out-of-core applications, whose datasets exceed the capacity of main memory, heavily rely on efficient I/O operations for high-performance. Several studies have shown that the I/O performance is far from the peak I/O performance in the majority of the applications~\cite{luu2015multiplatform,carns2011understanding}. MPI collective I/O addresses the challenge of the high-latency I/O operations with a generalized two-phase strategy for collective \texttt{read} and \texttt{write} accesses to shared files~\cite{thakur1999data}. Still, programmers need to use explicit I/O operations for interaction with the file system.

The previous work~\cite{mowry1996automatic} has pointed out that explicit I/O operations can have several disadvantages and proposes an approach from the compiler level. In fact,~\cite{jones2017unity} describes a new HPC-focused data storage abstraction that converges memory and storage. In this work, we provide a solution that extends the concept of one-sided communication model by enabling MPI windows with parallel I/O functionality. Hence, programmers maintain a familiar, unified programming interface.

\section{Conclusion And Future Work}
\label{6_Conclusion}

Computing nodes of next-generation supercomputers will include different memory and storage technologies. In this work, we proposed a novel use of MPI windows to hide the heterogeneity of the memory and storage subsystems by providing a single common programming interface for data movement across these layers. Our implementation, named MPI storage windows, is based on the memory-mapped file I/O mechanism of the OS. The approach transparently and efficiently integrates storage support into new and existing applications, without requiring any changes into the MPI standard. Moreover, it allows the definition of combined window allocations, that merge memory and storage under a unified virtual address space.

We evaluated MPI storage windows using two microbenchmarks and three different applications. We demonstrated that, while the approach features a performance degradation when compared to MPI memory windows, it can be effectively used for transparent checkpointing or exceeding the main memory capacity of the compute node. In most cases, the high-efficiency of MPI storage windows mostly relies on the fact that the page cache of the OS and the buffering of the parallel file system act as automatic caches for \texttt{read} and \texttt{write} operations on storage: in the same way that programmers do not necessarily handle explicit data movement in the processor caches, here programmers do not need to handle virtual memory management or buffering of the file system.

As future work, we plan to investigate the creation of a user-level memory-mapped I/O mechanism to provide full-control of storage allocations from the MPI implementation perspective. In addition, we plan to study the use of the {\ttfamily xpmem}~\cite{brightwell2011intra}\footnote{\url{https://github.com/hjelmn/xpmem}} Linux kernel module to map a remote storage window in the local virtual memory of an MPI process. Lastly, we plan to investigate the potential benefits of using Direct Access ({\ttfamily DAX})\footnote{\url{https://www.kernel.org/doc/Documentation/filesystems/dax.txt}}, an extension to the Linux kernel to map directly storage devices into virtual memory addresses. 

 \section*{Acknowledgement}
The experimental results were performed on resources provided by the Swedish National Infrastructure for Computing (SNIC) at PDC Centre for High Performance Computing (PDC-HPC).

The work was funded by the European Commission through the SAGE project (Grant agreement no. 671500 / http://www.sagestorage.eu). 


\bibliography{main}


\end{document}